\documentclass[11pt,a4paper]{article}

\usepackage[T1]{fontenc}
\usepackage{lmodern}
\usepackage{mathptmx}
\DeclareMathAlphabet{\mathcal}{OMS}{cmsy}{m}{n}
\usepackage[a4paper,margin=2.3cm,headheight=14pt]{geometry}
\usepackage{microtype}
\usepackage[numbers,sort&compress]{natbib}
\usepackage{etoolbox}
\usepackage{authblk}
\usepackage[font=small,labelfont=bf,labelsep=period]{caption}
\usepackage{titlesec}
\usepackage{fancyhdr}
\usepackage{graphicx}%
\usepackage{multirow}%
\usepackage{amsmath,amssymb,amsfonts}%
\usepackage{amsthm}%
\usepackage{mathrsfs}%
\usepackage{xcolor}%
\usepackage{textcomp}%
\usepackage{booktabs}%
\usepackage{algorithm}%
\usepackage{algorithmicx}%
\usepackage{algpseudocode}%
\usepackage{listings}%
\usepackage{subcaption}%
\usepackage{pifont}%
\usepackage{fontawesome5}%
\ifdefined\pdfmapfile\pdfmapfile{+fontawesome5.map}\fi
\usepackage{enumitem}%
\usepackage[export]{adjustbox}%
\usepackage{tikz}%
\usetikzlibrary{arrows.meta,positioning,shapes.geometric,shapes.symbols,fit,backgrounds,calc,shadows}

\newcommand{\tbd}[1]{\textcolor{red}{#1}}
\newcommand{\cmark}{\ding{51}}
\newcommand{\xmark}{\ding{55}}

\newcommand{\lpanel}[1]{\IfFileExists{figures/#1.pdf}{\includegraphics[max width=\linewidth]{figures/#1.pdf}}{\tbd{\scriptsize pending: \detokenize{#1}}}}
\newcommand{\panel}[1]{\IfFileExists{figures/#1.pdf}{\includegraphics[width=\linewidth]{figures/#1.pdf}}{\fbox{\parbox[c][2.2cm][c]{0.9\linewidth}{\centering\tbd{\scriptsize pending: \detokenize{#1}}}}}}

\definecolor{tbblue}{HTML}{2A78D6}
\definecolor{tborange}{HTML}{EB6834}
\definecolor{tbink}{HTML}{0B0B0B}
\definecolor{tbgray}{HTML}{52514E}
\definecolor{tblight}{HTML}{EAF2FC}

\theoremstyle{plain}%
\newtheorem{proposition}{Proposition}%
\theoremstyle{definition}%
\AtBeginEnvironment{table}{\centering}
\newcommand{\tabnote}[1]{\par\vspace{3pt}{\footnotesize\raggedright #1\par}}
\titleformat{\section}{\large\bfseries}{\thesection}{0.6em}{}
\titleformat{\subsection}{\normalsize\bfseries}{\thesubsection}{0.6em}{}
\titlespacing*{\section}{0pt}{1.6ex plus .4ex minus .2ex}{0.8ex}
\titlespacing*{\subsection}{0pt}{1.3ex plus .3ex minus .2ex}{0.6ex}

\usepackage{hyperref}
\hypersetup{colorlinks=true, linkcolor=tbblue, citecolor=tbblue, urlcolor=tbblue}

\title{\bfseries Trustworthy Blockchain-based Federated Learning for Electronic Health Records: Identity-Gated Participation against Sybil Attacks}
\author[1]{Rodrigo Tertulino\thanks{Corresponding author: \href{mailto:rodrigo.tertulino@ifrn.edu.br}{rodrigo.tertulino@ifrn.edu.br}}}
\author[1]{Ricardo Almeida}
\author[1]{Laercio Alencar}
\affil[1]{\small Federal Institute of Education, Science, and Technology of Rio Grande do Norte (IFRN), Natal, RN, Brazil}
\date{}

\begin{document}
\maketitle
\thispagestyle{empty}

\begin{abstract}
\noindent Federated learning lets hospitals train a shared clinical model without exchanging patient records, but its aggregation assumes that every participant is a distinct institution. In an open federation, an adversary can register many fake identities (a Sybil attack) and outweigh honest hospitals. Byzantine-robust aggregation mitigates such attacks only while the adversary is a minority. We ask whether binding participation to accountable institutional identity closes that gap, and at what cost. We present TBFL, a framework in which an accountable authority issues Verifiable Credentials to the Decentralized Identifiers of hospitals, and a smart-contract registry on a peer-to-peer ledger admits model updates only from credentialed senders. We evaluated a prototype on in-hospital mortality prediction with MIMIC-IV records split across simulated hospitals, comparing the identity gate with federated averaging, three robust aggregators, norm clipping, and a similarity-based defense. Robust aggregators held against a Sybil minority, and norm clipping only against random vectors; all failed when half the participants were Sybils trained on real records with flipped labels. The similarity-based defense stopped mutually similar Sybils but yielded to perturbed ones. The gate rejected all unauthorized submissions and maintained attack-free performance for every strategy up to 60\% Sybils. It did not stop credentialed hospitals that flipped their own labels, whereas the gate followed by the coordinate-wise median withstood both. On a local node, on-chain checks added about 1\% to training time; fees made training impractical on a public mainnet but affordable on a rollup. Identity gating complements robust aggregation rather than replacing it.

\medskip
\noindent\textbf{Keywords:} Identity-gated admission, Self-sovereign identity, Verifiable credentials, Smart contracts, Byzantine-robust aggregation, MIMIC-IV

\medskip
\noindent\textbf{Code and run logs:} \url{https://doi.org/10.5281/zenodo.23836753}
\end{abstract}

\section{Introduction}\label{sec:intro}

A hospital consortium that trains a mortality model via federated learning cannot tell whether ten incoming updates are from ten hospitals or from a single adversary with ten accounts. Federated learning (FL) is attractive for electronic health records (EHRs) because privacy regulation and institutional governance keep these records inside individual hospitals~\cite{shah2020}: participants exchange only model parameters, which an aggregator combines into a global model~\cite{mcmahan2017,rieke2020,kaissis2020}. FL, however, inherits the trust problems of every open distributed system, including the one in the opening sentence: a single entity presenting multiple identities. This Sybil attack~\cite{douceur2002} is especially damaging in FL because aggregation rules weight participants by their number or by the sample sizes they declare~\cite{fung2020,lian2023,samy2023}.

The standard algorithmic response is Byzantine-robust aggregation, such as Krum~\cite{blanchard2017} and the coordinate-wise median and trimmed mean~\cite{yin2018}. These rules bound the influence of a minority of outliers, but their guarantees require honest participants to be the majority. A Sybil adversary attacks exactly this assumption: identities are cheap, so the adversary can create as many as it needs to become the majority~\cite{fung2020}. Similarity-based defenses such as FoolsGold~\cite{fung2020} detect Sybils that behave identically, but an adversary that controls the content of its updates can diversify them. Blockchain-based FL records contributions on a ledger and often scores participants by their behavior~\cite{kim2020blockfl,lu2020,weng2021,qammar2023}. In such schemes, trust is established after the fact, new honest institutions start without reputation, and a patient adversary can accumulate reputation before attacking.

In healthcare, the set of legitimate participants is not anonymous: hospitals are licensed by public authorities. This makes a different defense possible, in which participation is bound to an institutional identity attested by an accountable issuer. The W3C standards for Decentralized Identifiers (DIDs)~\cite{w3c_did} and Verifiable Credentials (VCs)~\cite{w3c_vc} provide the building blocks, and self-sovereign identity (SSI) has been proposed for patient and provider identity in health systems~\cite{houtan2020,cocco2024,martinez2026}. Several works have combined SSI with FL to authenticate participants~\cite{geng2021didefed,haque2022,haque2024ssifl,zeydan2023,zeydan2024}. Their evaluations center on authentication and identity management. What remains under-examined is the security effect of identity gating on the model itself: how it compares with robust aggregation as the Sybil share grows, what it leaves unprotected, and what it costs when a smart contract on a peer-to-peer ledger, rather than an aggregator that checks certificates, enforces admission.

This paper addresses that question with an identity-gated FL framework, which we call TBFL, and an empirical study on clinical data. Its main contributions are:
\begin{enumerate}[label=\Roman*.]
\item \emph{System and threat model.} We define a system model, a threat model, and security goals for identity-gated FL. We prove that, under four stated assumptions, credential-bound admission excludes every uncredentialed identity from aggregation, and we state what it leaves unprotected (Sections~\ref{sec:system} and~\ref{sec:framework}).
\item \emph{Framework and prototype.} We design a ledger-based registry that admits model-update commitments only from credentialed senders, in two variants: an issuer-controlled allowlist and on-chain verification of issuer-signed EIP-712 credentials with expiry and revocation. We implement both, run federated training with FedProx through the allowlist, and show that both registries rejected every admission and update-binding attack (G1, G4) and the signed-credential registry rejected every credential attack (G2), including relays of old credentials after revocation, on a local node and on two public test networks (Sections~\ref{sec:framework} and~\ref{sec:res_security}).
\item \emph{Comparison with content-based defenses.} On in-hospital mortality prediction with MIMIC-IV records spread across ten heterogeneous simulated hospitals, we compare the identity gate with unprotected FedAvg, three Byzantine-robust aggregators, norm clipping and FoolsGold, and we combine the gate with the coordinate-wise median. The comparison covers Sybil shares ranging from minority to majority, with identical, adaptive, and label-flipping Sybils, as well as credentialed hospitals that poison their own updates (Section~\ref{sec:results}).
\item \emph{Measured cost.} We measure the gas and latency of admission and translate them into costs on a public ledger and on a rollup. We release the code, the cohort query, and all run logs (see Section~\ref{sec:results}).
\end{enumerate}

The study is organized around four research questions (RQs):
\begin{itemize}
\item \textbf{RQ1:} What predictive performance does the identity-gated federation reach on heterogeneous EHR data, relative to centralized and local-only training?
\item \textbf{RQ2:} How does identity gating compare with Byzantine-robust and similarity-based aggregation as the share of Sybil identities grows and as Sybils adapt their updates?
\item \textbf{RQ3:} Under which assumptions does identity gating prevent Sybil identities from influencing the global model, and does the implementation enforce it?
\item \textbf{RQ4:} What computational and monetary overhead does on-chain admission add, and how does it scale?
\end{itemize}

\section{Related work}\label{sec:related}

Table~\ref{tab:related} positions this work against representative studies. The comparison covers the technologies each study combines and how it is evaluated: whether identity or ledger overhead is measured, whether attacks are evaluated experimentally, whether the defense is compared with robust aggregation, and whether the setting is healthcare.

\begin{table}[t]
\caption{Overview of related studies. \cmark: present or supported; \xmark: absent or not evaluated. E1: identity or ledger overhead measured or analyzed; E2: attacks evaluated experimentally; E3: comparison with Byzantine-robust aggregation; E4: healthcare setting.}
\label{tab:related}
\footnotesize
\begin{tabular}{@{}lccccccccc@{}}
\toprule
& & \multicolumn{4}{c}{Technologies} & \multicolumn{4}{c}{Evaluation} \\
\cmidrule(lr){3-6}\cmidrule(l){7-10}
Study & Year & SSI & FL & Blockchain & Robust agg. & E1 & E2 & E3 & E4 \\
\midrule
Houtan et al.~\cite{houtan2020} & 2020 & \cmark & \xmark & \cmark & \xmark & \xmark & \xmark & \xmark & \cmark \\
Cocco and Tonelli~\cite{cocco2024} & 2024 & \cmark & \xmark & \cmark & \xmark & \xmark & \xmark & \xmark & \cmark \\
L\'opez Mart\'inez et al.~\cite{martinez2026} & 2026 & \cmark & \xmark & \cmark & \xmark & \cmark & \xmark & \xmark & \cmark \\
Kim et al.~\cite{kim2020blockfl} & 2020 & \xmark & \cmark & \cmark & \xmark & \cmark & \xmark & \xmark & \xmark \\
Fung et al.~\cite{fung2020} & 2020 & \xmark & \cmark & \xmark & \cmark & \xmark & \cmark & \cmark & \xmark \\
Cao et al.~\cite{cao2021fltrust} & 2021 & \xmark & \cmark & \xmark & \cmark & \xmark & \cmark & \cmark & \xmark \\
Geng et al.~\cite{geng2021didefed} & 2021 & \cmark & \cmark & \cmark & \xmark & \xmark & \xmark & \xmark & \cmark \\
Haque et al.~\cite{haque2022} & 2022 & \cmark & \cmark & \cmark & \xmark & \xmark & \xmark & \xmark & \xmark \\
Zeydan et al.~\cite{zeydan2024} & 2024 & \cmark & \cmark & \cmark & \xmark & \cmark & \xmark & \xmark & \xmark \\
Haque et al.~\cite{haque2024ssifl} & 2024 & \cmark & \cmark & \cmark & \xmark & \cmark & \xmark & \xmark & \cmark \\
This study & -- & \cmark & \cmark & \cmark & \cmark & \cmark & \cmark & \cmark & \cmark \\
\bottomrule
\end{tabular}
\end{table}

\paragraph{Poisoning and Sybil attacks in FL and their defenses.} In FL, model updates can be manipulated to degrade the global model (untargeted poisoning) or to implant behavior on chosen inputs (backdoors)~\cite{bagdasaryan2020,fang2020}. The main algorithmic response is Byzantine-robust aggregation, which replaces the weighted mean with a statistic that is robust to outliers. Krum selects the update with the smallest distance to its nearest neighbors~\cite{blanchard2017}. The coordinate-wise median and trimmed mean act on each parameter independently~\cite{yin2018}. Later work adapts these rules to decentralized topologies~\cite{fang2024,cajaraville2025} and refines the trimmed mean~\cite{wang2025trimmed}. All of them bound the fraction of malicious participants they can tolerate, and crafted updates can defeat them even below that bound~\cite{fang2020}. MESAS combines several statistical tests to detect adaptive attackers~\cite{krauss2023}.

FLTrust tolerates a malicious majority by anchoring trust in a small clean dataset held by the server, against which client updates are scored~\cite{cao2021fltrust}. In healthcare, this requires the server to hold representative clinical data, which is what the federation was set up to avoid. Douceur first described the Sybil attack in peer-to-peer systems: without an identity authority, a single entity can present arbitrarily many identities~\cite{douceur2002}. In FL, Fung et al.\ showed that Sybils submitting similar updates defeat robust aggregation. Their defense, FoolsGold, down-weights clients whose update histories are too similar~\cite{fung2020}. SPoiL demonstrated untargeted Sybil poisoning~\cite{lian2023}, and Samy and Girdzijauskas proposed further detection mechanisms~\cite{samy2023}. All of these defenses infer adversaries from the content or history of their updates, so an adversary that shapes its updates can evade the statistics in use. Privacy attacks such as gradient inversion~\cite{zhu2019dlg} are orthogonal to this paper. They are addressed by secure aggregation~\cite{bonawitz2017} and differential privacy~\cite{abadi2016}.

\paragraph{Blockchain-based federated learning.} Blockchain has been combined with FL to remove the single trusted aggregator, to record contributions tamper-evidently, and to reward participants~\cite{kim2020blockfl,lu2020,weng2021,nguyen2021,yang2023tfl}. Surveys catalog these designs~\cite{qammar2023,ning2024}, including the aggregation techniques used in blockchain-based FL for industrial IoT~\cite{shawkat2025}. Incentive mechanisms based on Shapley values~\cite{wang2022shapley} and architectures for accountability and fairness~\cite{lo2023} assess participants based on their output. The ledger then makes these assessments auditable, but admission is usually either open or controlled by a whitelist maintained by an operator. Reviews of blockchain in healthcare identify access control and identity management as open issues~\cite{hasselgren2020,holbl2018}.

\paragraph{Decentralized identity and its use in FL.} SSI has been proposed for patient identity~\cite{houtan2020}, for SSI--blockchain models of health services~\cite{cocco2024} and for role-based access to health records, enforced by smart contracts on Hyperledger Fabric whose throughput and latency were benchmarked~\cite{martinez2026}. Its privacy-by-design properties have also been analyzed systematically~\cite{naghmouchi2025}. Four works bring SSI into FL:
\begin{itemize}
\item DID-eFed uses DIDs and a smart contract to manage access to FL offered as a service among hospitals and research institutions~\cite{geng2021didefed}.
\item Haque et al.\ combined blockchain and SSI to authenticate FL participants~\cite{haque2022}.
\item SSI-FL, also by Haque et al., adds differential privacy and lets participants and issuers verify each other~\cite{haque2024ssifl}.
\item Zeydan et al.\ integrated blockchain-based SSI into vehicular FL and measured how credential operations scale~\cite{zeydan2023,zeydan2024}.
\end{itemize}
These works establish that SSI can authenticate FL participants, and their evaluations center on identity management, privacy, and overhead. As Table~\ref{tab:related} shows, none evaluate experimentally what identity gating does to model integrity under Sybil attack, or compare it with robust aggregation. These are the questions this paper addresses.

\section{Preliminaries}\label{sec:prelim}

TBFL combines four existing building blocks: federated optimization, which defines how updates are combined; robust and similarity-based aggregation, against which the gate is compared; self-sovereign identity, which supplies the credentials; and smart contracts, which enforce admission at a cost. Table~\ref{tab:notation} lists the notation.

\begin{table}[t]
\caption{Notation.}
\label{tab:notation}
\footnotesize
\begin{tabular}{@{}llll@{}}
\toprule
Symbol & Meaning & Symbol & Meaning \\
\midrule
$K$ & number of honest clients (hospitals) & $S$ & number of Sybil identities \\
$w^{t}$ & global model after round $t$ & $w_k^{t}$ & local model of client $k$ in round $t$ \\
$n_k$ & training samples of client $k$ & $F_k$ & local empirical loss of client $k$ \\
$\mu$ & FedProx proximal weight & $E, B$ & local epochs, batch size \\
$\mathcal{I}$ & trusted issuer & $\mathcal{P}$ & set of authorized addresses \\
$a_k$ & ledger address (DID) of client $k$ & $h_k^{t}$ & commitment $\mathrm{SHA256}(w_k^{t})$ \\
$\mathcal{S}_t$ & admitted updates in round $t$ & $\sigma$ & diversity of adaptive Sybil updates \\
$f$ & Krum's assumed number of attackers & $\beta$ & trimmed-mean trimming fraction \\
\bottomrule
\end{tabular}
\end{table}

\subsection{Federated learning and FedProx}

In round $t$, each client $k$ starts from the global model $w^{t-1}$ and minimizes its local loss. FedProx~\cite{li2020fedprox} adds a proximal term that limits client drift under heterogeneous data:
\begin{equation}
w_k^{t} = \arg\min_{w}\; F_k(w) + \frac{\mu}{2}\lVert w - w^{t-1}\rVert^2 .
\label{eq:fedprox}
\end{equation}
FedAvg~\cite{mcmahan2017} then forms the sample-weighted mean of the received updates, $w^{t} = \sum_{k} \frac{n_k}{\sum_j n_j}\, w_k^{t}$. Because the weights are proportional to the declared $n_k$, an identity that declares a large sample, or many identities, can dominate the mean.

\subsection{Byzantine-robust and similarity-based aggregation}

Robust aggregators replace the weighted mean with a statistic that bounds the influence of outliers. The coordinate-wise median takes, for each parameter $j$, $[w^t]_j=\operatorname{med}_k [w_k^t]_j$. The trimmed mean discards the fraction $\beta$ of the largest and smallest values for each parameter and then averages the remaining values~\cite{yin2018}. Krum assigns each update the score $\sum_{j\in\mathcal{N}_k}\lVert w_k^t - w_j^t\rVert^2$, where $\mathcal{N}_k$ are its $n-f-2$ nearest neighbors, and selects the update with the lowest score~\cite{blanchard2017}. All three assume that malicious updates are a minority; Krum explicitly requires $n>2f+2$. FoolsGold~\cite{fung2020} instead computes the pairwise cosine similarity of each client's accumulated update history, and assigns low aggregation weight to clients that are too similar to another client. We implement its core rule (pairwise similarity of accumulated updates, pardoning, and logit weighting) without the optional feature-importance weighting used in the original.

\subsection{Self-sovereign identity}

SSI rests on a trust triangle of issuer, holder, and verifier. A DID is a globally unique identifier controlled by a key pair. It is resolvable through a verifiable data registry, which can be a ledger; with the \texttt{did:ethr} method, the DID of an Ethereum account is derived from its address~\cite{w3c_did}. A VC is a set of claims about a subject DID, signed by an issuer~\cite{w3c_vc}. A verifier accepts a credential if the signature verifies against the issuer's key, the credential has not expired, and it has not been revoked. The issuer is accountable for the claims it signs. In healthcare, the natural issuers are licensing boards, health ministries, and consortium governance bodies.

\subsection{Smart contracts and transaction cost}

A smart contract is a program stored on a ledger whose state changes only through transactions. In the Ethereum Virtual Machine, each transaction pays for its computation and storage in \emph{gas}, at a price set by the network~\cite{wood2014}. Reverted transactions do not change state, and a client can detect a revert before broadcasting the transaction by estimating its gas. Rollups such as Arbitrum execute transactions off-chain and post compressed data to the main chain, thereby lowering the price per unit of gas~\cite{kalodner2018}. Permissioned ledgers such as Hyperledger Fabric charge no gas but require a consortium to operate the infrastructure~\cite{androulaki2018}.

\section{System overview}\label{sec:system}

Identity gating can only be judged against a stated adversary. This section specifies who participates in training, what the adversary can do, which assumptions the defense needs, and which goals it must meet; Section~\ref{sec:secanalysis} proves the goals follow from these assumptions.

\subsection{System model}\label{sec:sysmodel}

TBFL considers a consortium of hospitals that jointly train a clinical prediction model under the oversight of an accountable authority. As illustrated in Fig.~\ref{fig:overview}, the system comprises five entities:
\begin{itemize}
\item The \emph{trusted issuer} $\mathcal{I}$, typically a licensing board, a health ministry, or a consortium governance body, verifies institutions off-chain and issues them Verifiable Credentials.
\item Each \emph{hospital} $k$ holds a private dataset, a key pair whose address $a_k$ serves as its DID, and a credential issued by $\mathcal{I}$.
\item The \emph{registry contract}, deployed on a peer-to-peer ledger, records which addresses are authorized and admits, in every round, commitments to model updates.
\item The \emph{aggregator} distributes the global model and combines only the updates whose commitments were admitted.
\item The \emph{adversary} seeks to influence the global model by operating an arbitrary number of Sybil identities.
\end{itemize}

In every round, the interaction among these entities proceeds as follows. The aggregator distributes the current global model, and each hospital trains it on its local data. The hospital then commits the digest of its update to the registry and sends the update itself to the aggregator through an off-chain channel. Finally, the aggregator combines only those updates for which the registry has admitted commitments. Model parameters are never written on-chain: only the digest $h_k^t=\mathrm{SHA256}(w_k^t)$~\cite{fips180} of the float32 parameters is recorded, encoded in the prototype as a 64-character hexadecimal string. Consequently, the ledger cost remains independent of model size, and the confidentiality of the parameters does not depend on the ledger.

\begin{figure}[t]
\centering
\begin{tikzpicture}[
  >={Stealth[length=1.8mm]}, font=\footnotesize,
  box/.style={draw=tbgray, rounded corners=2pt, align=center, inner sep=3pt, fill=white, minimum height=10mm, text width=#1},
  box/.default=2.9cm,
  chainbox/.style={box=3.1cm, draw=tbblue, fill=tblight, line width=0.7pt},
  advbox/.style={box=3.4cm, draw=tborange, line width=0.7pt},
  lbl/.style={font=\scriptsize, fill=white, inner sep=1.2pt},
  ok/.style={->, tbink, line width=0.6pt},
  bad/.style={->, tborange, line width=0.6pt, dashed}]
\node[box, double copy shadow={shadow xshift=2pt, shadow yshift=-2pt, draw=tbgray, fill=white}] (h) at (0,0)
  {{\Large\color{tbgray}\faHospital}\\[2pt] \textbf{Hospitals} $1,\dots,K$\\ DID $a_k$, VC, local EHR};
\node[chainbox] (reg) at (4.95,1.9) {{\Large\color{tbblue}\faCubes}\\[2pt] \textbf{Registry contract}\\ peer-to-peer ledger};
\node[box] (agg) at (9.85,0) {{\Large\color{tbgray}\faServer}\\[2pt] \textbf{Aggregator}\\ combines admitted $w_k^t$};
\node[box] (iss) at (0,3.55) {{\Large\color{tbgray}\faLandmark}\\[2pt] \textbf{Trusted issuer} $\mathcal{I}$\\ licensing authority};
\node[advbox] (adv) at (4.95,-2.45) {{\Large\color{tborange}\faUserSecret}\\[2pt] \textbf{Adversary}\\ Sybil identities $s_1,\dots,s_S$};
\draw[ok] (iss) -- node[lbl, right] {(1) issue VC \faIdCard[regular]} (h);
\draw[ok] (iss.east) -| node[lbl, pos=0.4, above] {(2) authorize $a_k$ after VC check} (reg.north);
\draw[ok] (h.north east) -- node[lbl, above left=1pt] {(3) commit $h_k^t$} (reg.south west);
\draw[ok] (reg.south east) -- node[lbl, above right=1pt] {(4) admitted $h_k^t$} (agg.north west);
\draw[ok] ([yshift=1.5mm]h.east) -- node[lbl, above, pos=0.22] {(5) $w_k^t$, off-chain} ([yshift=1.5mm]agg.west);
\draw[ok] ([yshift=-1.5mm]agg.west) -- node[lbl, below, pos=0.78] {(6) global model $w^t$} ([yshift=-1.5mm]h.east);
\draw[bad] (adv.north) -- node[lbl, right, pos=0.3, text=tborange] {forged VC, $h_s^t$: \faTimesCircle\ \textbf{revert}} (reg.south);
\draw[bad] (adv.east) -| node[lbl, pos=0.25, below, text=tborange] {$w_s^t$: \faTimesCircle\ \textbf{discarded}} (agg.south);
\end{tikzpicture}
\caption{System model of TBFL. Solid arrows are the honest protocol; dashed arrows are Sybil attempts, which the registry reverts and the aggregator discards. Only the digest $h_k^t=\mathrm{SHA256}(w_k^t)$ is written on-chain.}
\label{fig:overview}
\end{figure}

\subsection{Threat model}\label{sec:threatmodel}

The threat model follows the classical formulation of the Sybil attack in peer-to-peer systems, in which a single entity presents multiple identities in order to gain disproportionate influence~\cite{douceur2002}. In federated learning, such influence translates directly into aggregation weight, since aggregation rules weight participants by their number or by the sample sizes they declare. This subsection specifies the capabilities attributed to the adversary, the assumptions on which the framework relies, and the threats that are deliberately excluded from the analysis.

\paragraph{Adversary capabilities.} We consider a strong adversary that fully controls the content of every Sybil update. Specifically, the adversary can:
\begin{itemize}
\item create any number of key pairs and ledger accounts;
\item obtain the current global model;
\item send arbitrary update vectors to the aggregator and arbitrary transactions to the ledger, including calls to the registry's authorization function;
\item coordinate all Sybil identities and choose their updates: identical across Sybils, individually perturbed to evade similarity-based detection, or trained on data from the honest distribution with inverted labels, so that they resemble honest updates in scale.
\end{itemize}

\paragraph{Assumptions.} The security of TBFL rests on the following four assumptions. We state them explicitly because each one is invoked in the security analysis of Section~\ref{sec:secanalysis}:
\begin{enumerate}
\item[(A1)] The issuer is honest and its signing key is not compromised.
\item[(A2)] Signatures and hash functions are secure.
\item[(A3)] The ledger provides integrity and eventual inclusion of valid transactions.
\item[(A4)] The aggregator follows the protocol and aggregates exactly the updates whose commitments were admitted in the current round. Relaxing it would require verifiable aggregation or a trusted execution environment, which we leave for future work.
\end{enumerate}

\paragraph{Out of scope.} Finally, the following threats lie outside the scope of this paper. Although relevant in practice, they are addressed by complementary mechanisms, such as robust aggregation, secure aggregation and differential privacy, which can be layered on top of TBFL (see Section~\ref{sec:discussion}):
\begin{itemize}
\item credentialed insiders, that is, legitimate hospitals that poison their own updates;
\item compromise or collusion of the issuer;
\item inference attacks on updates;
\item denial of service against the ledger.
\end{itemize}

\subsection{Security goals}\label{sec:goals}

Building on the system and threat models above, TBFL is designed to achieve four security goals. Goals G1 and G2 concern \emph{who} may contribute to the global model, whereas G3 and G4 concern the auditability and integrity of the contributions that are admitted:
\begin{enumerate}
\item[(G1)] \emph{Credential-bound admission:} only updates from issuer-authorized identities contribute to the global model.
\item[(G2)] \emph{Unforgeable credentials:} an identity cannot obtain authorization with a forged credential, a credential issued to another subject, or an expired, revoked, or superseded credential, including by re-relaying a credential that was once valid.
\item[(G3)] \emph{Accountability:} every authorization and every admitted commitment is recorded tamper-evidently, together with its sender.
\item[(G4)] \emph{Update binding:} the aggregator uses only updates that match the commitment admitted for their sender in the current round.
\end{enumerate}
These goals concern the admission and integrity of contributions, not their content. They do not require that admitted updates be benign, which is the reason why credentialed insiders are excluded from the threat model and why identity gating is complementary to robust aggregation. Experiment~5 (Section~\ref{sec:design}) measures what happens when this exclusion is violated.

\section{The TBFL framework}\label{sec:framework}

TBFL implements the security goals through two mechanisms: a credential check upon hospital admission and a commitment check in every round.

\subsection{Credential issuance and admission}\label{sec:admission}

As illustrated in Fig.~\ref{fig:protocol}, the protocol comprises an onboarding phase, executed once per institution, and a training phase, repeated in every round. During onboarding, a hospital presents its license and accreditation to the issuer (step~1), which performs due diligence (step~2). If the issuer accepts the documentation, it signs an EIP-712 typed credential \texttt{HospitalCredential(subject, expiresAt, issuedAt)} that binds the hospital's address to an expiry date and an issuance time (step~3). Any party can relay the credential to the registry (step~4): \texttt{authorize\allowbreak Worker} recovers the signer with \texttt{ecrecover}, rejects the credential unless the signer is $\mathcal{I}$, the credential is unexpired, and it was issued after the last credential accepted for, or the last revocation of, the same subject; only then does it mark the subject as authorized (step~5). A Sybil that relays a forged credential is rejected at this point (attempt~A). The issuer can later call \texttt{revoke\allowbreak Credential}, which voids all credentials issued to the subject up to that point. Because relayed credentials are public on the ledger, the issuance check prevents anyone from re-relaying an old credential to undo a revocation or roll back a renewed expiry; a credential issued after the revocation re-authorizes the hospital.

In addition to this signed-credential registry, we implement a simpler allowlist variant, \texttt{FLRegistry}, which skips the on-chain signature check: the issuer verifies the credential off-chain and calls \texttt{authorize\allowbreak Worker} itself. Once authorized, a hospital takes part in the training phase. In every round, it receives the global model (step~6), trains locally (step~7), and commits $h_k^t$ through \texttt{submit\allowbreak Update} (step~8). The registry emits an event for an authorized sender (step~9) and reverts otherwise, so a Sybil's commitment never enters the ledger (attempt~B). The hospital then sends $w_k^t$ to the aggregator off-chain (step~10). The aggregator reads the admitted commitments (step~11) and aggregates only the updates that match them (step~12); an update without an admitted commitment, such as a Sybil's, is discarded (attempt~C). Algorithms~\ref{alg:contract} and~\ref{alg:fl} formalize, respectively, the registry contract and one training round. TBFL is the admission gate; the rule that aggregates the admitted updates is a parameter. In the experiments, \emph{TBFL} denotes the gate followed by FedAvg and \emph{TBFL + median} the gate followed by the coordinate-wise median, which Section~\ref{sec:discussion} recommends as the default. The contract checks expiry and revocation when each commitment is mined, so a revocation takes effect for every commitment included after it, including later commitments in the same round; a commitment admitted before the revocation remains part of that round's aggregate.

\begin{figure}[t]
\centering
\begin{tikzpicture}[
  font=\scriptsize, >={Stealth[length=1.6mm]},
  head/.style={draw=tbgray, rounded corners=2pt, minimum width=20mm, minimum height=6mm, align=center, fill=white, font=\footnotesize},
  msg/.style={->, tbink, line width=0.55pt},
  bad/.style={->, tborange, line width=0.55pt, dashed},
  act/.style={draw=tbgray, fill=white, rounded corners=1pt, inner sep=2pt, align=center, text width=#1},
  act/.default=26mm,
  frag/.style={draw=tbgray!70, rounded corners=2pt},
  tab/.style={draw=tbgray!70, fill=white, font=\scriptsize\itshape, inner sep=2pt, anchor=north west},
  lbl/.style={above, inner sep=1.5pt},
  stp/.style={circle, fill=tbblue, text=white, font=\tiny\bfseries, inner sep=0pt, minimum size=3.1mm},
  atk/.style={circle, fill=tborange, text=white, font=\tiny\bfseries, inner sep=0pt, minimum size=3.1mm}]
\def\xI{1.2}\def\xH{3.95}\def\xR{6.7}\def\xA{9.45}\def\xS{12.0}
\node[head] at (\xI,0) {{\color{tbgray}\faLandmark}\ Issuer $\mathcal{I}$};
\node[head] at (\xH,0) {{\color{tbgray}\faHospital}\ Hospital $k$};
\node[head, draw=tbblue, fill=tblight, line width=0.7pt] at (\xR,0) {{\color{tbblue}\faCubes}\ Registry};
\node[head] at (\xA,0) {{\color{tbgray}\faServer}\ Aggregator};
\node[head, draw=tborange, line width=0.7pt] at (\xS,0) {{\color{tborange}\faUserSecret}\ Sybil $s$};
\foreach \x in {\xI,\xH,\xR,\xA,\xS} \draw[tbgray, densely dotted] (\x,-0.32) -- (\x,-10.5);
\draw[frag] (0.0,-0.55) rectangle (13.05,-4.80); \node[tab] at (0.0,-0.55) {onboarding};
\draw[msg] (\xH,-1.30) -- node[lbl, pos=0.4] {license, accreditation} (\xI,-1.30);
\node[stp] at (\xH-0.25,-1.48) {1};
\node[act=20mm] (n2) at (\xI,-1.80) {due diligence};
\node[stp] at (n2.north west) {2};
\draw[msg] (\xI,-2.50) -- node[lbl] {\faIdCard[regular]\ VC $=\mathrm{Sig}_{\mathcal{I}}(a_k,\mathit{exp},\mathit{iss})$} (\xH,-2.50);
\node[stp] at (\xI+0.25,-2.68) {3};
\draw[msg] (\xH,-3.10) -- node[lbl] {\texttt{authorize\allowbreak Worker}(VC)} (\xR,-3.10);
\node[stp] at (\xH+0.25,-3.28) {4};
\node[act=34mm] (n5) at (\xR,-3.70) {signer $=\mathcal{I}$, unexpired,\\ newer $\Rightarrow a_k\in\mathcal{P}$};
\node[stp] at (n5.north west) {5};
\draw[bad] (\xS,-4.55) -- node[lbl, text=tborange, pos=0.3] {forged VC: \faTimesCircle\ \textbf{revert}} (\xR,-4.55);
\node[atk] at (\xS+0.28,-4.55) {A};
\draw[frag] (0.0,-5.00) rectangle (13.05,-10.35); \node[tab] at (0.0,-5.00) {round $t$};
\draw[msg] (\xA,-5.60) -- node[lbl] {$w^{t-1}$} (\xH,-5.60);
\node[stp] at (\xA-0.25,-5.78) {6};
\node[act=36mm] (n7) at (\xH,-6.10) {local FedProx training};
\node[stp] at (n7.north west) {7};
\draw[msg] (\xH,-6.80) -- node[lbl] {\texttt{submit\allowbreak Update}$(h_k^t)$} (\xR,-6.80);
\node[stp] at (\xH+0.25,-6.98) {8};
\node[act=31mm] (n9) at (\xR,-7.30) {$a_k\in\mathcal{P}$: emit event};
\node[stp] at (n9.north west) {9};
\draw[msg] (\xH,-8.00) -- node[lbl, pos=0.25] {$w_k^t$ (off-chain)} (\xA,-8.00);
\node[stp] at (\xH+0.25,-8.18) {10};
\draw[bad] (\xS,-8.60) -- node[lbl, text=tborange, pos=0.3] {$h_s^t$: \faTimesCircle\ \textbf{revert}} (\xR,-8.60);
\node[atk] at (\xS+0.28,-8.60) {B};
\draw[msg] (\xR,-9.20) -- node[lbl] {admitted $h_k^t$} (\xA,-9.20);
\node[stp] at (\xR+0.25,-9.38) {11};
\draw[bad] (\xS,-9.20) -- node[lbl, text=tborange] {$w_s^t$: \faTimesCircle\ \textbf{discarded}} (\xA,-9.20);
\node[atk] at (\xS+0.28,-9.20) {C};
\node[act=40mm] (n12) at (\xA,-9.9) {aggregate the $w_k^t$ that match $h_k^t$};
\node[stp, anchor=east] at ([xshift=-0.6mm]n12.west) {12};
\end{tikzpicture}
\caption{Admission protocol and one training round with the signed-credential registry. Blue numbers are the steps referred to in Section~\ref{sec:admission}; orange letters are Sybil attempts (dashed arrows). Boxes on a lifeline are local actions. In the allowlist variant, the issuer itself calls \texttt{authorize\allowbreak Worker} after verifying the credential off-chain.}
\label{fig:protocol}
\end{figure}

\begin{algorithm}[t]
\caption{Registry contract (on-chain), signed-credential variant. The FL experiments use the allowlist variant, in which only the issuer may call \textsc{authorizeWorker}}\label{alg:contract}
\begin{algorithmic}[1]
\State \textbf{state:} $\mathcal{I}$ (deployer); $\mathit{expiry}:\text{address}\to\mathbb{N}$; $\mathit{revoked}:\text{address}\to\text{bool}$; $\mathit{minIss}:\text{address}\to\mathbb{N}$
\Function{authorizeWorker}{$a, \mathit{exp}, \mathit{iss}, \mathit{sig}$}
  \State $d \gets \mathrm{EIP712}(\textsc{HospitalCredential}(a,\mathit{exp},\mathit{iss}))$
  \If{$\mathrm{ecrecover}(d,\mathit{sig})\neq\mathcal{I}$ \textbf{or} $\mathit{exp}\le\mathit{now}$ \textbf{or} $\mathit{iss}>\mathit{now}$ \textbf{or} $\mathit{iss}\le\mathit{minIss}[a]$} \State \textbf{revert} \EndIf
  \State $\mathit{minIss}[a]\gets\mathit{iss}$; $\mathit{expiry}[a]\gets\mathit{exp}$; $\mathit{revoked}[a]\gets\mathsf{false}$
\EndFunction
\Function{revokeCredential}{$a$} \Comment{only $\mathcal{I}$}
  \State $\mathit{revoked}[a]\gets\mathsf{true}$; $\mathit{minIss}[a]\gets\max(\mathit{minIss}[a],\mathit{now})$
\EndFunction
\Function{submitUpdate}{$h$}
  \If{$\mathit{expiry}[\mathtt{msg.sender}]\le\mathit{now}$ \textbf{or} $\mathit{revoked}[\mathtt{msg.sender}]$} \State \textbf{revert} \EndIf
  \State record $h$; \textbf{emit} \textsc{ModelUpdated}$(h, \mathtt{msg.sender})$
\EndFunction
\end{algorithmic}
\end{algorithm}

\begin{algorithm}[t]
\caption{Identity-gated FedProx round $t$ (off-chain)}\label{alg:fl}
\begin{algorithmic}[1]
\Require global model $w^{t-1}$, proximal weight $\mu$
\State $\mathcal{S}_t\gets\emptyset$
\For{each participant $k$ that sends an update $w_k^t$ with declared size $n_k$}
  \State honest clients compute $w_k^t$ by Eq.~(\ref{eq:fedprox}) with $E$ epochs of SGD
  \State $h_k\gets\mathrm{SHA256}(w_k^t)$; send \Call{submitUpdate}{$h_k$} from address $a_k$
  \If{the transaction is included, not reverted, sent by $a_k$, and its \textsc{ModelUpdated} event carries $\mathrm{SHA256}(w_k^t)$} \State $\mathcal{S}_t\gets\mathcal{S}_t\cup\{k\}$
  \Else \State discard $w_k^t$ and log the rejection \EndIf
\EndFor
\State $w^t\gets\mathrm{Aggregate}(\{w_k^t\}_{k\in\mathcal{S}_t})$ \Comment{FedAvg or a robust rule (Section~\ref{sec:discussion})}
\end{algorithmic}
\end{algorithm}

\subsection{Security analysis}\label{sec:secanalysis}

We now show that the protocol described above achieves the security goals of Section~\ref{sec:goals} under the assumptions of Section~\ref{sec:threatmodel}. The following proposition summarizes the result.

\begin{proposition}\label{prop:gate}
Under (A1)--(A4), TBFL satisfies G1--G4: no update sent by an identity without a valid credential issued by $\mathcal{I}$ contributes to the global model.
\end{proposition}

\begin{proof}
The argument covers the four goals in turn.

\emph{G2.} \texttt{authorize\allowbreak Worker} marks $a$ as authorized only if \texttt{ecrecover} returns $\mathcal{I}$ for the EIP-712 digest of $(a,\mathit{exp},\mathit{iss})$ and $\mathit{exp}$ lies in the future. By (A2), producing such a signature without $\mathcal{I}$'s key is infeasible. The digest also binds the subject $a$, so a credential issued to another subject does not verify. The digest binds the issuance time as well, and the contract accepts only a credential issued after the last accepted credential and the last revocation of $a$, so re-relaying an old credential reverts. Revocation and expiry are checked on every submission.

\emph{G1.} \texttt{submit\allowbreak Update} reverts for unauthorized senders, and by (A3) a transaction cannot be included with a forged sender. By (A4) the aggregator uses only admitted updates, so an unauthorized identity never enters $\mathcal{S}_t$.

\emph{G3.} Every state change is a ledger transaction whose sender is recorded, and every admitted commitment emits an event. By (A3), these records are tamper-evident. The allowlist used in training emits no event on authorization, so its authorizations are audited through the issuer's transactions; \texttt{FLRegistry\allowbreak Events} adds events that name the authorized subject.

\emph{G4.} The aggregator reads the admitted commitment $h_k^t$ from the event emitted by the transaction of $a_k$ and keeps the update received off-chain only if $\mathrm{SHA256}(w_k^t)=h_k^t$. By (A2), no other vector matches the admitted commitment.
\end{proof}

Proposition~\ref{prop:gate} follows from access control. It is not a learning-theoretic guarantee, and it says nothing about the quality of admitted updates. Its strength rests on (A1): the cost of a Sybil identity becomes the cost of deceiving or corrupting the issuer, not the cost of generating a key pair. Section~\ref{sec:res_security} tests each case of the proof against the deployed contracts. Table~\ref{tab:scope} separates the components evaluated in this paper from the design elements left for future work.

\begin{table}[t]
\caption{Scope of the evaluated prototype.}
\label{tab:scope}
\footnotesize
\begin{tabular}{@{}lll@{}}
\toprule
Component & Status & Evaluated in \\
\midrule
Issuer allowlist (\texttt{FLRegistry}) & Implemented & all FL runs \\
Allowlist with audit events (\texttt{FLRegistry\allowbreak Events}) & Implemented & registry benchmark \\
EIP-712 credentials, expiry, revocation (\texttt{FLRegistry\allowbreak ZK}) & Implemented & registry benchmark \\
SHA-256 commitments; aggregation of admitted updates & Implemented & all FL runs \\
Aggregator check of update against commitment (G4) & Implemented & benchmark; 41 FL runs$^{\mathrm{a}}$ \\
Gate followed by a robust rule (median) & Implemented & Exp.~2 and~5 \\
Zero-knowledge membership proofs (Groth16) & Released & not evaluated \\
Off-chain parameter storage (e.g.\ IPFS~\cite{benet2014}) & Design only & -- \\
Public test networks (Sepolia, Arbitrum Sepolia) & Deployed & testnet benchmark \\
Permissioned network & Not deployed & -- \\
\bottomrule
\end{tabular}
\tabnote{$^{\mathrm{a}}$29 runs with Sybils (Experiments~2 to~4) and 12 with credentialed insiders (Experiment~5); the 12 earliest runs of Experiment~2 predate the check.}
\end{table}

\section{Experiments and results}\label{sec:results}

The evaluation asks four questions in turn: how well the federation predicts without attack (RQ1, Section~\ref{sec:res_perf}), how the gate compares with content-based defenses under Sybil attack (RQ2, Section~\ref{sec:res_sybil}), whether the deployed registry enforces admission (RQ3, Section~\ref{sec:res_security}) and what admission costs (RQ4, Section~\ref{sec:res_cost}). Sections~\ref{sec:testbed}--\ref{sec:metrics} first describe the setup.

\subsection{Testbed}\label{sec:testbed}

All experiments ran on a single commodity machine without GPU acceleration, which suffices for the model size studied. Table~\ref{tab:testbed} summarizes the tools, environment, deployments, tests and metrics.

\begin{table}[t]
\caption{Summary of the testbed.}
\label{tab:testbed}
\footnotesize\setlength{\tabcolsep}{4pt}
\begin{tabular}{@{}p{0.17\linewidth}p{0.165\linewidth}p{0.2\linewidth}p{0.21\linewidth}p{0.165\linewidth}@{}}
\toprule
Tools & Environment & Deployment & Tests & Metrics \\
\midrule
\raggedright $\bullet$ PyTorch 2.15 \newline
$\bullet$ scikit-learn 1.9 \newline
$\bullet$ imbalanced-learn 0.14 \newline
$\bullet$ web3.py 8.0 \newline
$\bullet$ Hardhat 2.22 \newline
$\bullet$ Solidity 0.8.24
&
\raggedright $\bullet$ 10-core 64-bit ARM CPU \newline
$\bullet$ 16~GB RAM \newline
$\bullet$ no GPU \newline
$\bullet$ Python 3.13
&
\raggedright $\bullet$ local Hardhat node (EVM, automine) \newline
$\bullet$ Ethereum Sepolia \newline
$\bullet$ Arbitrum Sepolia (rollup) \newline
$\bullet$ 3 registry variants \newline
$\bullet$ up to 10 concurrent runs
&
\raggedright $\bullet$ FL Experiments 1--5 (3--5 seeds) \newline
$\bullet$ registry benchmark (10 clients, 20 rounds) \newline
$\bullet$ testnet benchmark (5 clients, 10 rounds per registry)
&
\raggedright $\bullet$ AUC-ROC, AUPRC \newline
$\bullet$ Brier score \newline
$\bullet$ precision, recall \newline
$\bullet$ accepted / attempted \newline
$\bullet$ gas, latency, US\$ \tabularnewline
\bottomrule
\end{tabular}
\end{table}

Up to ten runs executed concurrently, each on its own block of ledger accounts. Every run logs its metrics for each round and every transaction it sends: sender, round, outcome, gas used, and latency from submission to receipt. Code development was assisted by an AI tool; the authors reviewed the code and take full responsibility for all content.

\subsection{Dataset}

We use MIMIC-IV v3.1~\cite{johnson2023mimic}, accessed through PhysioNet~\cite{goldberger2000} under its credentialed data use agreement. The task is in-hospital mortality (\texttt{hospital\_expire\_flag}), predicted from seven variables:
\begin{itemize}
\item the patient's anchor age, sex, race and insurance;
\item the first unit in the admission's transfer history, which can be the emergency department or a ward;
\item the number of diagnosis codes and of procedure codes recorded for the admission.
\end{itemize}
One-hot encoding expands these variables to 32 inputs. Every admission of a patient with an anchor age of at least 18 is included, and no admission is excluded afterward. Two variables have missing values, and none is imputed: insurance is missing for 9{,}355 admissions (1.7\%), which form their own category, and the first unit for 4, which join the less frequent units in an ``other'' category; unknown or declined race forms its own category. Table~\ref{tab:dataset} summarizes the data and the released SQL query reproduces the cohort.

The cohort is \emph{not} restricted to intensive-care stays or to first admissions. The diagnosis and procedure counts are complete only at discharge. The task is therefore retrospective risk stratification of a completed admission, used here as a realistic, heavily imbalanced tabular benchmark for the security study. Reporting follows TRIPOD~\cite{collins2015tripod} where it applies.

\begin{table}[t]
\caption{Statistics of the dataset and of the federation (seed 0).}
\label{tab:dataset}
\footnotesize
\begin{tabular}{@{}lr@{}}
\toprule
Measure & Value \\
\midrule
Admissions (adult patients) & 546{,}028 \\
Distinct patients & 223{,}452 \\
In-hospital deaths & 11{,}801 (2.16\%) \\
Raw variables / encoded inputs & 7 / 32 \\
Clients and partition & 10, label-wise Dirichlet ($\alpha=0.5$) \\
Admissions per client (min--max) & 8{,}653--105{,}576 \\
In-hospital mortality per client (min--max) & 0.11\%--10.5\% \\
Global test set (union of local 20\% splits) & 109{,}210 admissions, 2{,}360 deaths \\
\bottomrule
\end{tabular}
\end{table}

\begin{figure}[t]
\centering
\begin{subfigure}[t]{0.48\textwidth}\centering
\panel{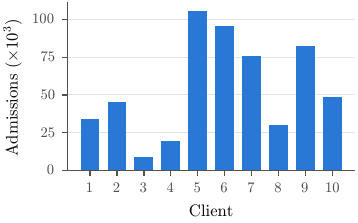}
\caption{Admissions per client}\label{fig:data_a}
\end{subfigure}\hfill
\begin{subfigure}[t]{0.48\textwidth}\centering
\panel{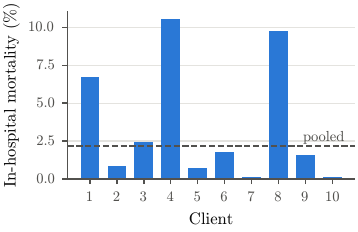}
\caption{In-hospital mortality per client}\label{fig:data_b}
\end{subfigure}
\caption{Heterogeneity of the federation produced by the Dirichlet partition (seed 0). The dashed line marks the pooled mortality.}
\label{fig:data}
\end{figure}

Admissions are split across $K=10$ clients by a label-wise Dirichlet partition~\cite{hsu2019} with concentration $\alpha=0.5$. The partition is redrawn until every client has at least 2{,}000 admissions and 50 deaths; for seed~0, the smallest client has 8{,}653 admissions. Figure~\ref{fig:data} shows the resulting heterogeneity: client sizes differ by an order of magnitude, and mortality ranges from 0.11\% to 10.5\%. Each client then makes a stratified 80/20 train/test split. Age and the two counts are standardized with training-set statistics. SMOTETomek~\cite{batista2004} is applied \emph{only} to each local training split, after the split, so no resampled record enters a test set. The global test set is the union of the local test sets and keeps the natural prevalence. Partitions are regenerated for each random seed.

\subsection{Parameters and experimental design}\label{sec:design}

Table~\ref{tab:params} lists all parameters. They were fixed in advance and are identical across conditions; no parameter was tuned on the test set.

\begin{table}[t]
\caption{Parameters of the experiments.}
\label{tab:params}
\footnotesize
\begin{tabular}{@{}lll@{}}
\toprule
Component & Parameter & Value \\
\midrule
\multirow{4}{*}{Model} & architecture & MLP 32--64--32--1, ReLU, dropout 0.2 (4{,}225 parameters) \\
& loss & binary cross-entropy + proximal term, $\mu=0.01$ \\
& optimizer & SGD, learning rate 0.01, momentum 0.9, weight decay $10^{-5}$ \\
& local training & $E=3$ epochs, batch size $B=32$ \\
\midrule
\multirow{3}{*}{Federation} & clients & $K=10$, all participate in every round \\
& rounds & $R=100$ (Exp.~1 and early stopping), $R=50$ (Exp.~2--5) \\
& seeds & 5 (Exp.~1, Exp.~2, Exp.~3 with $\sigma=1$); 3 (other conditions) \\
\midrule
\multirow{6}{*}{Attack} & start & round 10 \\
& Sybil identities & $S\in\{2,5,10,15\}$ (17\%, 33\%, 50\%, 60\% of participants) \\
& update & $v\sim\mathcal{N}(0,I)$; adaptive: $v+\sigma\varepsilon_s$, $\varepsilon_s\sim\mathcal{N}(0,I)$, \\
& & $\sigma\in\{0.1,0.3,1\}$; label flipping: 5{,}000 records, inverted labels \\
& declared size & mean honest training-set size \\
& insiders & $k\in\{1,2\}$ largest clients, own data, inverted labels \\
\midrule
\multirow{6}{*}{Defenses} & robust, under attack & median; trimmed mean $\beta=\min(S/n,\,0.49)$; \\
& & Krum $f=\min(S,\lfloor(n-3)/2\rfloor)$ \\
& robust, no attack & trimmed mean $\beta=0.1$; Krum $f=1$ \\
& FoolsGold & cosine similarity of update histories, pardoning, logit weighting \\
& norm clipping & deltas clipped to the median norm of the round, then FedAvg \\
& gate + median & TBFL admission, then coordinate-wise median \\
\midrule
\multirow{2}{*}{Ledger} & node & Hardhat 2.22 (EVM, automine), Solidity 0.8.24 \\
& cost model & 20~gwei (Ethereum L1), 0.1~gwei (Arbitrum), ETH $=$ US\$3{,}000 \\
\bottomrule
\end{tabular}
\end{table}

The study comprises five experiments and four auxiliary analyses:
\begin{itemize}
\item \textbf{Experiment 1 (no attack).} TBFL runs for 100 rounds with five seeds. It is compared against two baselines that use the same model, optimizer, and features: \emph{centralized} training on the union of clients' training data, and \emph{local-only} training on each client's own data. Both baselines train for 30 epochs. The training budgets therefore differ: in 100 rounds of three local epochs, each client passes over its own data 300 times, whereas the centralized model passes over the pooled data 30 times; the early-stopping analysis below shows how much of the federated result depends on the number of rounds.
\item \textbf{Experiment 2 (identical Sybils).} From round 10, $S$ Sybil identities submit the same random vector and declare the mean honest sample size, so FedAvg gives them the weight of average clients. We compare TBFL with unprotected FedAvg, the coordinate-wise median, the trimmed mean, Krum, FoolsGold, and norm clipping, and we combine the gate with the median. Norm clipping scales every update delta down to the median delta norm of the round, a common defense against oversized updates. The robust rules receive the true $S$, an oracle choice that favors them, capped where the rule requires it: Krum needs $n>2f+2$, so $f=8$ for $S=10$ and $f=11$ for $S=15$, and the trimmed mean must keep at least one value per coordinate.
\item \textbf{Experiment 3 (adaptive Sybils).} Each Sybil adds independent noise to the common vector. With $\sigma=1$, the expected cosine similarity between Sybils drops from 1 to 0.5, thereby targeting similarity-based detection. For $S=10$ we also vary $\sigma\in\{0.1,0.3\}$ for the median, Krum, and FoolsGold, and we test the gate with $S=15$. Under TBFL, each Sybil first attempts to authorize itself, then submits a commitment in each round.
\item \textbf{Experiment 4 (label-flipping Sybils).} Random vectors are easy to recognize by their norm. In this experiment, each Sybil instead behaves like a plausible client: in every round, it trains the current global model using the same local procedure as honest clients, on 5{,}000 records drawn from an honest client's training data with inverted labels. This assumes an adversary with access to data from the same distribution. We test $S\in\{5,10\}$ against all defenses, and $S=15$ against the gate.
\item \textbf{Experiment 5 (credentialed insiders).} The gate admits every credentialed hospital, so it cannot stop a hospital that poisons its own updates. Here, without Sybils, the $k\in\{1,2\}$ largest clients, which carry the most FedAvg weight, train on their own data with inverted labels from round 10 and declare their true size. We compare the gate followed by FedAvg with the gate followed by the median.
\item \textbf{Early-stopping analysis.} To quantify the effect of reporting the last round, we repeat FedAvg without attack for 100 rounds with 10\% of each client's training split held out, before resampling, as a validation set. We compare the round with the best validation AUC with the last round. Without attack, the gate admits every client, so this applies to TBFL as well.
\item \textbf{Patient-level sensitivity analysis.} Because the 546{,}028 admissions belong to 223{,}452 patients, the same patient can appear in training and test data. We therefore repeat centralized training using the split made by patient rather than by admission.
\item \textbf{Registry benchmark.} We benchmark the three registry variants with ten clients over 20 rounds, including every attack case of Proposition~\ref{prop:gate}.
\item \textbf{Testnet benchmark.} We deploy the allowlist and signed-credential registries on Ethereum Sepolia and on Arbitrum Sepolia. Five clients submit commitments for ten rounds per registry, and the attack cases are repeated so that confirmation latency and gas are measured on public peer-to-peer networks.
\end{itemize}

\subsection{Evaluation metrics}\label{sec:metrics}

Because only 2.16\% of admissions end in death, we report:
\begin{itemize}
\item the area under the ROC curve (AUC-ROC);
\item the area under the precision--recall curve (AUPRC), whose chance level equals the prevalence~\cite{saito2015};
\item the Brier score, as a measure of probability accuracy~\cite{vancalster2019}; a constant prediction of the prevalence scores $0.0216\times0.9784\approx0.021$;
\item precision and recall at the default threshold of 0.5; F1 and the false-negative rate are in the released logs.
\end{itemize}
Results are given as mean $\pm$ standard deviation over seeds at the last round, without selecting the best round, except in the early-stopping analysis. With three to five seeds, we report differences between means together with the standard deviation across seeds, rather than significance tests. Because most differences between the better defenses are smaller than the standard deviation across seeds, the tables do not mark a best value.

\subsection{Predictive performance without attack (RQ1)}\label{sec:res_perf}

Table~\ref{tab:perf} compares the identity-gated federation with the baselines. After 100 rounds, TBFL reached an AUC-ROC of $0.869\pm0.031$ and an AUPRC of $0.167\pm0.049$, which is 7.7 times the prevalence. Centralized training on the same training data reached $0.876\pm0.006$ and $0.171\pm0.015$. Federation therefore cost 0.007 in AUC, less than the standard deviation across seeds, while every record stayed at its client; the price was greater variation across seeds. Models trained by each client alone reached $0.852\pm0.008$ on the global test set.

The lower block of Table~\ref{tab:perf} addresses the repeated-patient concern. With a single global split and resampling, splitting by admission or by patient gave the same AUC ($0.895\pm0.000$ and $0.895\pm0.003$) and almost the same AUPRC ($0.218$ and $0.221$). Repeated patients therefore do not inflate discrimination in this task. These rows also suggest that resampling on the pooled training set yields higher discrimination than resampling inside each skewed client (0.895 versus 0.876). The two configurations, however, also differ in their split and test set.

\begin{figure}[t]
\centering
\begin{subfigure}[t]{0.32\textwidth}\centering
\panel{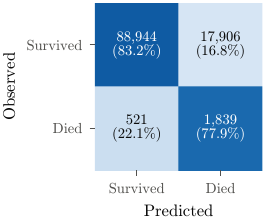}\caption{Confusion matrix (seed 0)}\end{subfigure}\hfill
\begin{subfigure}[t]{0.32\textwidth}\centering
\panel{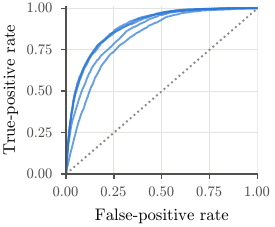}\caption{ROC curves}\end{subfigure}\hfill
\begin{subfigure}[t]{0.32\textwidth}\centering
\panel{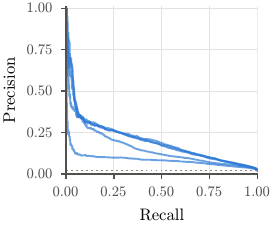}\caption{Precision--recall curves}\end{subfigure}
\caption{Final TBFL models after round 100 on the global test set (109{,}210 admissions, 2{,}360 deaths). In (b) and (c), each line is one seed and the dotted line is a random classifier.}
\label{fig:clin}
\end{figure}

The low precision at the 0.5 threshold ($0.093\pm0.006$) is expected. Training on balanced data shifts scores toward the positive class, and at the natural prevalence even a specific classifier produces many false positives per true positive. For seed~0 the model identified 1{,}839 of the 2{,}360 deaths, with 17{,}906 false alarms among 106{,}850 survivors (Fig.~\ref{fig:clin}a). The same shift affects calibration. All models in Table~\ref{tab:perf} have a Brier score between 0.099 and 0.136, worse than the 0.021 of a constant prediction of the prevalence, so their outputs rank admissions well but are not probabilities. Any clinical use would require threshold selection and recalibration~\cite{vancalster2019}.

\begin{table}[t]
\caption{Predictive performance without attack (mean $\pm$ SD over five seeds; threshold 0.5). Upper block: global test set for the federation (prevalence 2.16\%); baselines are trained for 30 epochs. A constant prediction of the prevalence has a Brier score of 0.021.}
\label{tab:perf}
\footnotesize\setlength{\tabcolsep}{3.2pt}
\begin{tabular}{@{}lccccc@{}}
\toprule
Model & AUC-ROC & AUPRC & Brier & Precision & Recall \\
\midrule
TBFL (federated) & $0.869\pm0.031$ & $0.167\pm0.049$ & $0.099\pm0.031$ & $0.093\pm0.006$ & $0.642\pm0.236$ \\
Centralized, same data & $0.876\pm0.006$ & $0.171\pm0.015$ & $0.122\pm0.046$ & $0.092\pm0.035$ & $0.726\pm0.150$ \\
Local only$^{\mathrm{a}}$ & $0.852\pm0.008$ & $0.144\pm0.006$ & -- & $0.090\pm0.003$ & $0.578\pm0.040$ \\
\midrule
\multicolumn{6}{@{}l}{\emph{Centralized, sensitivity to repeated patients}$^{\mathrm{b}}$} \\
Admission split & $0.895\pm0.000$ & $0.218\pm0.006$ & $0.128\pm0.006$ & $0.085\pm0.003$ & $0.819\pm0.010$ \\
Patient split & $0.895\pm0.003$ & $0.221\pm0.011$ & $0.136\pm0.004$ & $0.080\pm0.004$ & $0.838\pm0.013$ \\
\bottomrule
\end{tabular}
\tabnote{$^{\mathrm{a}}$Mean over clients of each client's model evaluated on the global test set; the Brier score was not recorded for these models.}
\tabnote{$^{\mathrm{b}}$Single global 80/20 split with SMOTETomek on the pooled training set; test sets of 109{,}206 (admission split) and 109{,}409 admissions (patient split), so these rows are not directly comparable with the upper block.}
\end{table}

\begin{figure}[t]
\centering
\lpanel{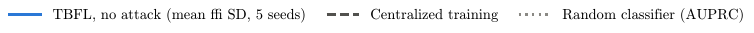}\\[2pt]
\begin{subfigure}[t]{0.32\textwidth}\centering
\panel{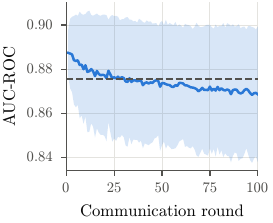}\caption{AUC-ROC}\end{subfigure}\hfill
\begin{subfigure}[t]{0.32\textwidth}\centering
\panel{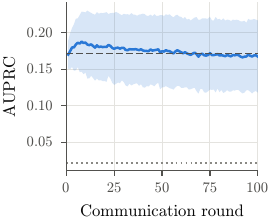}\caption{AUPRC}\end{subfigure}\hfill
\begin{subfigure}[t]{0.32\textwidth}\centering
\panel{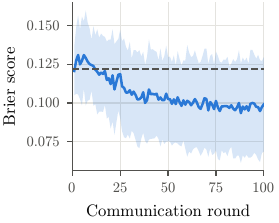}\caption{Brier score}\end{subfigure}
\caption{Learning curves of TBFL without attack on the global test set (mean $\pm$ SD over five seeds).}
\label{fig:conv}
\end{figure}

Figure~\ref{fig:conv} shows the learning curves. Discrimination peaked early: each seed reached its best AUC (0.869--0.899) within the first ten rounds and then declined slowly. For three seeds, the decline was small, with AUCs ranging from 0.885 to 0.891. For seeds 2 and 3, it was pronounced, and these two seeds account for most of the dispersion in Table~\ref{tab:perf}. We attribute the drift to local overfitting on resampled data, amplified by partitions where a few clients account for most deaths. The early-stopping analysis confirms this. With a held-out validation set, the round with the best validation AUC fell between rounds 1 and 5 for all five seeds. At that round, the test AUC was $0.889\pm0.014$ and the AUPRC $0.180\pm0.028$, against $0.870\pm0.028$ and $0.167\pm0.052$ at round 100 of the same runs. The selected models were more sensitive (recall $0.79$ versus $0.61$) but less well calibrated (Brier $0.122$ versus $0.093$). For FedAvg, validation-based stopping therefore closes most of the gap to the robust rules seen in Section~\ref{sec:res_sybil}; we did not apply it to the robust rules, which drift less. The other tables keep the last round, so that no reported number involves model selection. Figure~\ref{fig:clin} shows the confusion matrix and the ROC and precision--recall curves of the final models.

\begin{figure}[t]
\centering
\lpanel{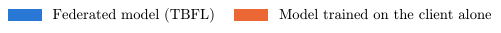}\\[2pt]
\begin{subfigure}[t]{0.48\textwidth}\centering
\panel{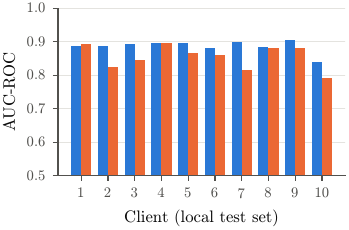}\caption{AUC-ROC}\end{subfigure}\hfill
\begin{subfigure}[t]{0.48\textwidth}\centering
\panel{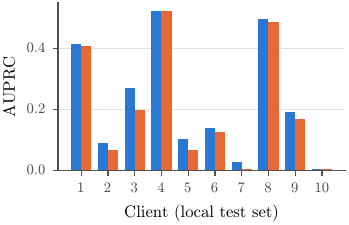}\caption{AUPRC}\end{subfigure}
\caption{Performance on each client's local test set of the federated model and of a model trained only on that client's data (seed 0).}
\label{fig:clients}
\end{figure}

Figure~\ref{fig:clients} compares, client by client, the federated model with a model trained on the client alone, both evaluated on the client's local test set. On these local test sets, the federated model achieved a mean AUC of $0.868\pm0.030$, compared with $0.850\pm0.016$ for local-only models. It improved the AUC for 33 of the 50 client--seed pairs (one-sided sign test $p=0.016$, treating pairs as independent although clients of one seed share a federation), with a mean gain of 0.017.

The gain was concentrated among clients with low mortality: its correlations were $-0.47$ with local mortality and $0.42$ with test-set size. For seed~0, clients 2, 7, and 10, whose mortality ranged from 0.11\% to 0.88\%, gained between 0.05 and 0.08 in AUC. Clients 4 and 8, whose mortality was close to 10\%, gained almost nothing. AUPRC did not change on average ($-0.003$), because it is dominated by local prevalence. Hospitals with few deaths, which have the fewest positive examples to learn from, are thus the ones that benefit most from joining the federation.

\subsection{Robustness to Sybil attacks (RQ2)}\label{sec:res_sybil}

\begin{figure}[t]
\centering
\includegraphics[width=\linewidth]{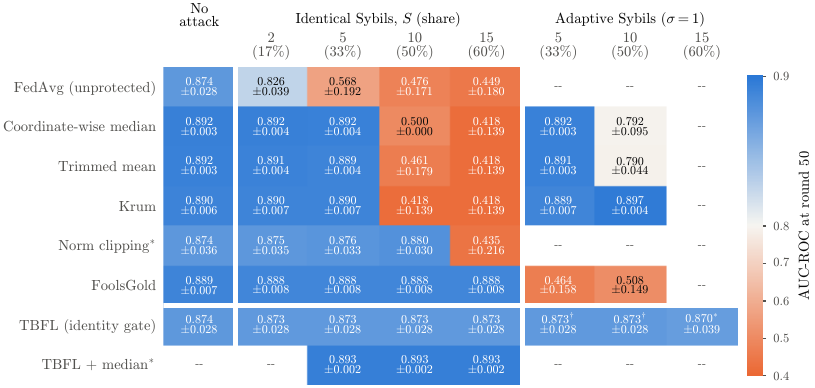}
\caption{AUC-ROC at round 50 under a Sybil attack starting at round 10, by defense and attack condition (mean $\pm$ SD over five seeds unless marked). $S$ Sybil identities join 10 honest clients; robust aggregators receive the true $S$, capped as in Table~\ref{tab:params}. The color scale is centered at 0.8: orange cells indicate defenses that failed, while blue cells indicate defenses that maintained their attack-free accuracy. --: condition not run. $^{*}$Three seeds. $^{\dagger}$The gate rejects Sybils by identity before their update content is observed, so the adaptive attack yields the same admitted set as the identical attack.}
\label{fig:heat}
\end{figure}

\begin{figure}[t]
\centering
\lpanel{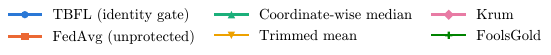}\\[2pt]
\begin{subfigure}[t]{0.48\textwidth}\centering
\panel{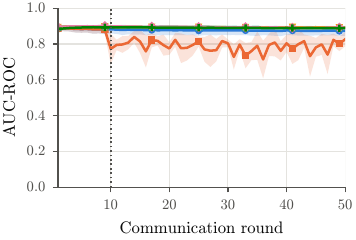}\caption{$S=2$ (17\%)}\end{subfigure}\hfill
\begin{subfigure}[t]{0.48\textwidth}\centering
\panel{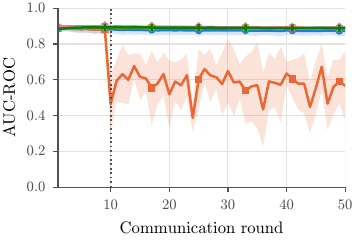}\caption{$S=5$ (33\%)}\end{subfigure}\\[4pt]
\begin{subfigure}[t]{0.48\textwidth}\centering
\panel{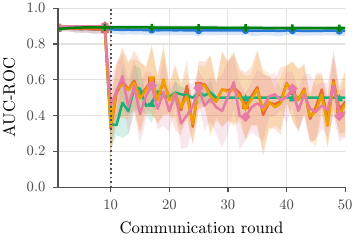}\caption{$S=10$ (50\%)}\end{subfigure}\hfill
\begin{subfigure}[t]{0.48\textwidth}\centering
\panel{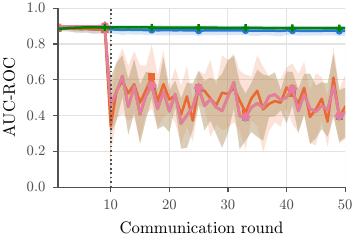}\caption{$S=15$ (60\%)}\end{subfigure}
\caption{Global AUC-ROC per round under identical Sybils joining at round 10 (dotted line). Mean $\pm$ SD over five seeds.}
\label{fig:sybil}
\end{figure}

\begin{figure}[t]
\centering
\lpanel{sybil_legend}\\[2pt]
\begin{subfigure}[t]{0.48\textwidth}\centering
\panel{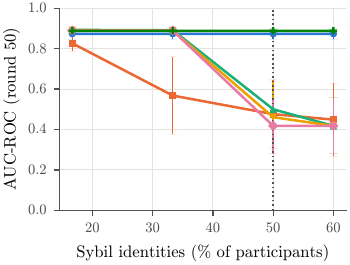}\caption{Identical Sybils, by Sybil share}\end{subfigure}\hfill
\begin{subfigure}[t]{0.48\textwidth}\centering
\panel{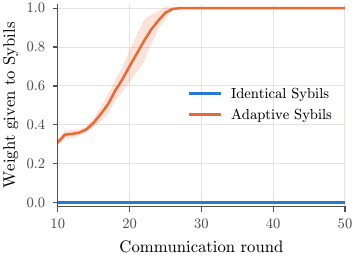}\caption{Weight FoolsGold gives to Sybils ($S=5$)}\end{subfigure}\\[4pt]
\lpanel{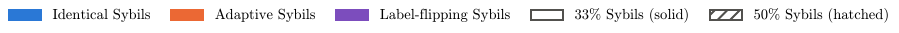}\\[2pt]
\begin{subfigure}[t]{0.48\textwidth}\centering
\panel{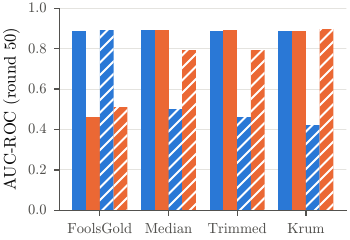}\caption{Content-based defenses under identical and adaptive Sybils}\end{subfigure}\hfill
\begin{subfigure}[t]{0.48\textwidth}\centering
\panel{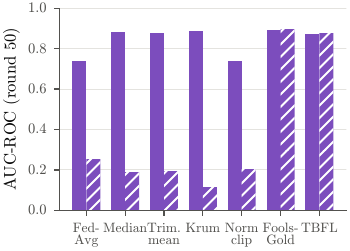}\caption{All defenses under label-flipping Sybils}\end{subfigure}
\caption{Content-based defenses against identity-based gating. (a) AUC-ROC at round 50 against the share of identical Sybils (legend at top). The dotted line marks 50\%. (b) Aggregation weight that FoolsGold assigns to adaptive and identical Sybils. (c) AUC-ROC at round 50 for each content-based defense under identical and adaptive Sybils. (d) AUC-ROC at round 50 under label-flipping Sybils (three seeds). Bars: mean over seeds (five in (c)).}
\label{fig:frac}
\end{figure}

Does any content-based defense survive once Sybils become the majority, and does the identity gate change the answer? Figs.~\ref{fig:heat}, \ref{fig:sybil} and~\ref{fig:frac} report identical and adaptive Sybils (Experiments 2 and 3), Table~\ref{tab:variants} the $\sigma$ sweep and label-flipping Sybils (Experiment~4), and Table~\ref{tab:insiders} credentialed insiders (Experiment~5). We take the defenses in turn, from unprotected FedAvg to the gate, and end with the attack that the gate does not cover.

\paragraph{Unprotected FedAvg.} FedAvg degraded with two Sybils (17\%, AUC $0.826\pm0.039$) and collapsed with five or more ($0.568\pm0.192$ at 33\%), because one identical noise vector carrying a third or more of the aggregation weight dominates the mean.

\paragraph{Robust aggregators.} The robust rules behaved as their theory predicts. With Sybils in the minority, they matched their attack-free reference and were more accurate and more stable than TBFL's FedAvg ($0.892\pm0.004$ for the median versus $0.873\pm0.028$ at 33\%), because the median damps the slow drift of seeds 2 and 3 described in Section~\ref{sec:res_perf}. At 50\% and 60\% all three failed. The coordinated Sybils formed the majority, or the tightest cluster, of identical vectors, so the median adopted their values and Krum selected one of them. With ten Sybils, the median returned a constant-score model (AUC 0.500), and Krum and the trimmed mean fell below chance. The repeated entries of $0.418\pm0.139$ are not a coincidence: whenever a rule outputs the Sybil vector itself, the global model is the same random vector for every rule, and its AUC on a given seed is fixed.

\paragraph{Norm clipping.} Clipping every delta to the median norm of the round neutralized the oversized random vectors while they were a minority, up to $S=10$ ($0.880\pm0.030$). With 20 participants, the implementation uses the lower of the two middle norms, which remains the largest honest norm when the Sybil norms are all larger. With 15 Sybils, the median norm became a Sybil norm, and the defense failed ($0.435\pm0.216$).

\paragraph{FoolsGold.} Against identical Sybils, FoolsGold was the only content-based defense that held at every share. Their update histories had a cosine similarity of 1, so it assigned them a weight of 0 from the first attack round on and kept the AUC at $0.888\pm0.008$ for every Sybil share, including the 60\% majority. This equals its attack-free value ($0.889\pm0.007$), as expected when the Sybils receive no weight. The adaptive attack reversed this result. Once each Sybil added independent noise, the Sybils became less similar to one another than the honest hospitals, whose updates share the same clinical signal. FoolsGold therefore down-weighted the honest clients instead. With five Sybils, their share of the aggregation weight rose from about one third at round 10 to one by round 26 (Fig.~\ref{fig:frac}b), and AUC fell to $0.464\pm0.158$. This failure follows from the defense's premise that Sybils resemble each other more than honest clients do.

\paragraph{Robust aggregators under adaptive Sybils.} The adaptive attack did not affect the robust rules with a Sybil minority. With ten Sybils and $\sigma=1$, the median and the trimmed mean degraded ($0.792\pm0.095$ and $0.790\pm0.044$) but did not collapse, because the independent noise partly cancels coordinate by coordinate. Krum held ($0.897\pm0.004$): the noisy Sybil updates lie far from each other and from the honest cluster, so none of them scored best. The $\sigma$ sweep in Table~\ref{tab:variants} shows that no single noise level defeats every rule. At $\sigma=0.1$ and $0.3$, the Sybils still fixed the median at a constant score (0.500), while Krum and FoolsGold already held; at $\sigma=0$ Krum failed, and at $\sigma=1$ FoolsGold failed. An adversary that knows the defense in use can therefore choose a $\sigma$ that defeats it.

\paragraph{Label-flipping Sybils.} Sybils that trained on real records with inverted labels produced updates through the same training procedure as honest clients, so norm clipping did not separate them (Table~\ref{tab:variants} and Fig.~\ref{fig:frac}d). With five Sybils, the median, the trimmed mean, and Krum held ($0.883\pm0.005$, $0.879\pm0.005$ and $0.889\pm0.003$), while FedAvg and norm clipping lost about 0.13 in AUC. With ten Sybils, every robust rule and norm clipping fell below chance (0.11 to 0.25): the flipped updates formed a coherent majority that pulled the model toward the inverted labels. FoolsGold held at both sizes ($0.892\pm0.004$ at $S=10$) because Sybils that start from the same global model and optimize the same inverted objective have similar updates; they accounted for less than 4\% of the weight in every round. Its success, therefore, rests on the same premise of similarity that the adaptive attack violates.

\paragraph{Identity gate.} Under TBFL, every Sybil self-authorization reverted, and every Sybil commitment was rejected: $S$ rejections per round, none accepted, for identical, adaptive and label-flipping Sybils alike. Performance matched the attack-free run at every Sybil share ($0.873\pm0.028$ against $0.874\pm0.028$). With 15 adaptive or 15 label-flipping Sybils, run for three seeds, each seed ended within 0.01 of its attack-free AUC ($0.870\pm0.039$ and $0.869\pm0.038$); the dispersion comes from seed~2, whose attack-free AUC is 0.819. Combining the gate with the coordinate-wise median gave $0.893\pm0.002$ at 5, 10 and 15 Sybils, the median's attack-free accuracy, whereas the median alone collapsed at 50\%. By Proposition~\ref{prop:gate}, these outcomes hold by construction; the experiments confirm that the implementation enforces them and that the gate composes with a robust rule.

\paragraph{Credentialed insiders.} The gate gave no protection once the attacker held a credential (Table~\ref{tab:insiders}). When the largest client flipped its labels, the gate followed by FedAvg ended at 0.886 and 0.874 for two seeds but at 0.480 for the third, whose attack-free AUC at round 50 is 0.830. When the two largest clients flipped theirs, it ended at 0.810, 0.219, and 0.204. These clients have the largest training sets, so FedAvg gives their updates the largest weights. The gate followed by the median was unaffected: $0.894\pm0.002$ with one insider and $0.893\pm0.001$ with two, the same as against Sybils. One or two insiders are a minority of the ten admitted hospitals, which is the condition the median needs.

\begin{table}[t]
\caption{AUC-ROC and AUPRC at round 50 with credentialed insiders (Experiment~5; mean $\pm$ SD over three seeds). From round 10, the $k$ largest clients train on their own data with inverted labels; there are no Sybils. Attack-free references are in Fig.~\ref{fig:heat}.}
\label{tab:insiders}
\footnotesize
\begin{tabular}{@{}lcccc@{}}
\toprule
& \multicolumn{2}{c}{$k=1$} & \multicolumn{2}{c}{$k=2$} \\
\cmidrule(lr){2-3}\cmidrule(l){4-5}
Configuration & AUC-ROC & AUPRC & AUC-ROC & AUPRC \\
\midrule
TBFL (gate + FedAvg) & $0.746\pm0.231$ & $0.103\pm0.078$ & $0.411\pm0.346$ & $0.031\pm0.032$ \\
TBFL + median & $0.894\pm0.002$ & $0.205\pm0.002$ & $0.893\pm0.001$ & $0.200\pm0.006$ \\
\bottomrule
\end{tabular}
\end{table}

\begin{table}[t]
\caption{AUC-ROC at round 50 under further Sybil strategies (mean $\pm$ SD over three seeds). Left: adaptive Sybils with smaller noise, $S=10$ ($\sigma=0$ and $\sigma=1$ are in Fig.~\ref{fig:heat}). Right: label-flipping Sybils (Experiment~4).}
\label{tab:variants}
\footnotesize\setlength{\tabcolsep}{4pt}
\begin{tabular}{@{}lcccc@{}}
\toprule
& \multicolumn{2}{c}{Adaptive, $S=10$} & \multicolumn{2}{c}{Label flipping} \\
\cmidrule(lr){2-3}\cmidrule(l){4-5}
Defense & $\sigma=0.1$ & $\sigma=0.3$ & $S=5$ (33\%) & $S=10$ (50\%) \\
\midrule
FedAvg (unprotected) & -- & -- & $0.741\pm0.156$ & $0.250\pm0.024$ \\
Coordinate-wise median & $0.500\pm0.000$ & $0.500\pm0.000$ & $0.883\pm0.005$ & $0.186\pm0.039$ \\
Trimmed mean & -- & -- & $0.879\pm0.005$ & $0.191\pm0.046$ \\
Krum & $0.893\pm0.003$ & $0.896\pm0.003$ & $0.889\pm0.003$ & $0.113\pm0.014$ \\
Norm clipping & -- & -- & $0.737\pm0.139$ & $0.202\pm0.031$ \\
FoolsGold & $0.893\pm0.004$ & $0.893\pm0.004$ & $0.893\pm0.003$ & $0.892\pm0.004$ \\
TBFL (identity gate) & -- & -- & $0.872\pm0.036$ & $0.873\pm0.032$ \\
\bottomrule
\end{tabular}
\end{table}

\subsection{Security tests of the registry (RQ3)}\label{sec:res_security}

Table~\ref{tab:sectests} lists every attack case of Proposition~\ref{prop:gate} and its outcome on the deployed contracts. None of the attempts were accepted, neither in the benchmark nor during federated training. In particular, re-relaying the credential of a revoked hospital, which is public on the ledger, reverted; only a credential issued after the revocation re-authorized it, after which its next commitment was accepted. All of them reverted during gas estimation, were never mined, and therefore cost the honest participants nothing. The same property means that rejected attempts also cost the adversary nothing, so the gate offers no economic deterrent to flooding the ledger's RPC endpoints; we treat this as denial-of-service, which is out of scope. The lower rows of the benchmark block test G4 off-chain: the aggregator accepted an update that matched its admitted commitment and rejected both a substituted update and an update attributed to a different sender. A validation run with this check enabled reproduced the recorded learning trajectory exactly, and the 41 TBFL runs that ran with the check, 29 with Sybils in Experiments~2 to~4 and 12 with insiders in Experiment~5, found no mismatch; the 12 earliest runs of Experiment~2 predate the check. Three of the 31{,}500 commitments that credentialed clients sent during training (0.010\%) were not recorded, and the affected client missed that round's aggregation. The senders were authorized, and their other commitments in the same runs were accepted, so we attribute these failures to transient errors of the local node; the logs do not record the error message. All three occurred in the first batch of runs, in which ten runs shared the node concurrently; the 18 later runs, with at most seven concurrent runs, lost none.

\begin{table}[t]
\caption{Attack attempts against the registry and their outcomes (accepted/attempted). Upper block: registry benchmark with ten clients over 20 rounds; lower block: all 41 TBFL runs with Sybils in Experiments~2 to~4, of which 29 also ran the G4 check; the 12 earliest runs of Experiment~2 predate it.}
\label{tab:sectests}
\footnotesize
\begin{tabular}{@{}lllc@{}}
\toprule
Attempt & Goal & Registry & Accepted \\
\midrule
Self-authorization by a non-issuer & G1 & both allowlists & 0 / 2 \\
Commitment from an unauthorized address & G1 & all three variants & 0 / 60 \\
Credential signed by a key other than $\mathcal{I}$ & G2 & signed credential & 0 / 1 \\
Valid credential of another subject (replay) & G2 & signed credential & 0 / 1 \\
Expired credential & G2 & signed credential & 0 / 1 \\
Commitment after revocation & G2 & signed credential & 0 / 1 \\
Accepted credential relayed again & G2 & signed credential & 0 / 1 \\
Old credential relayed after revocation & G2 & signed credential & 0 / 1 \\
Commitment after that relay & G2 & signed credential & 0 / 1 \\
Update substituted after its commitment & G4 & all three variants & 0 / 3 \\
Update attributed to another sender & G4 & all three variants & 0 / 3 \\
\midrule
Sybil self-authorization during FL training & G1 & allowlist & 0 / 385 \\
Sybil commitments during FL training & G1 & allowlist & 0 / 15{,}785 \\
\bottomrule
\end{tabular}
\end{table}

\subsection{Efficiency analysis (RQ4)}\label{sec:res_cost}

\begin{figure}[t]
\centering
\lpanel{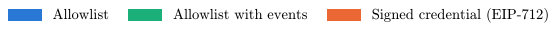}\\[2pt]
\begin{subfigure}[t]{0.32\textwidth}\centering
\panel{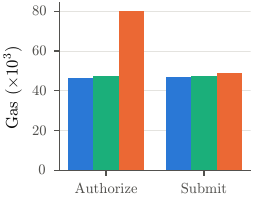}\caption{Gas per operation}\end{subfigure}\hfill
\begin{subfigure}[t]{0.32\textwidth}\centering
\panel{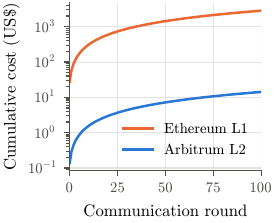}\caption{Cumulative cost}\end{subfigure}\hfill
\begin{subfigure}[t]{0.32\textwidth}\centering
\panel{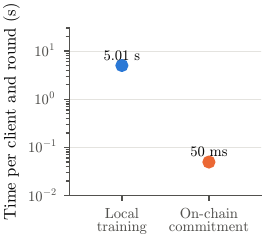}\caption{Time per client and round}\end{subfigure}
\caption{Efficiency of on-chain admission. (a) Median gas per operation for the three registry variants. (b) Cumulative cost of the allowlist over 100 rounds on Ethereum L1 and on Arbitrum (log scale). (c) Local training time versus commitment latency on the local node (log scale).}
\label{fig:cost}
\end{figure}

On-chain admission is affordable only if its gas and latency are small compared to training. Table~\ref{tab:cost} and Fig.~\ref{fig:cost} report both on the local node, and Table~\ref{tab:testnet} reports on public test networks.

\paragraph{Allowlist used in the FL experiments.} The first commitment of a run cost about 2.5 times the median (117{,}765 gas), because it initializes the storage slots of the task record; later commitments only overwrite them. Over 100 rounds, 10 clients used 47.4 million gas units, and cumulative gas usage grew linearly with the number of rounds ($R^2=1.000$).

\paragraph{Signed-credential registry.} Verifying the EIP-712 signature and recording the issuance time on-chain raised the cost of authorization by 73\%, a one-off cost per credential, and the expiry and revocation checks raised the cost of a commitment by 4\% (Table~\ref{tab:cost}). Gas does not depend on model size, because only a digest is recorded. Rejected transactions cost honest participants nothing.

\paragraph{Monetary cost.} At 20~gwei and US\$3{,}000 per ETH, a ten-client round costs about US\$28 on Ethereum mainnet, which is too expensive for routine training. On an Arbitrum rollup, it costs about US\$0.14, excluding layer-1 data fees. Because the gas per round is fixed, the cost scales linearly with the gas and ETH prices, both of which vary by more than an order of magnitude over time. At the same ETH price, a mainnet round would cost about US\$1.41 at 1~gwei and US\$7.03 at 5~gwei; Table~\ref{tab:cost} can be rescaled to any other assumption.

\paragraph{Latency.} On the local node, a commitment took $49.6\pm1.8$~ms, against $5.01\pm0.07$~s of local training per client and round, about 1\% overhead. This figure covers signing, the RPC round trip, and EVM execution, but not block production, which the testnet benchmark below measures.

\paragraph{Public test networks.} Table~\ref{tab:testnet} reports the deployments on Arbitrum Sepolia and Ethereum Sepolia. On both networks, every attack attempt reverted, and all 100 commitments per network matched their events, so the admission and binding results of the local benchmark carry over. Arbitrum Sepolia confirmed a single transaction in a median of 1.2~s and reproduced the local gas figures, up to a few units of calldata gas that depend on the signature bytes, at a gas price of 0.02~gwei, below the 0.1~gwei assumed in Table~\ref{tab:cost}. Ethereum Sepolia needed a median of 10.6~s, about one slot, which is twice the local training time of a round; on a layer-1 network, commitments must therefore be submitted asynchronously while the next round trains. Our round times are upper bounds, because one machine issued the five commitments in sequence through a public RPC endpoint; with each hospital submitting from its own node, a round's ledger time approaches the single-transaction latency.

Sepolia also revealed a change in the gas schedule. Since October 2026, Sepolia has been running the Glamsterdam upgrade, which reprices the creation of, and access to, contract state~\cite{ef2026glamsterdam}. Under this schedule, a commitment costs 62{,}273 gas (33\% more than on the local node), an authorization costs 128{,}353 gas (2.8 times more; 250{,}062 for a signed credential, 3.1 times more), and a deployment costs about seven times more, because authorizations and deployments create new state. If mainnet adopts the same schedule, a ten-client round of the allowlist would cost about US\$37 at 20~gwei and US\$1.9 at 1~gwei, instead of US\$28 and US\$1.4. Rollups, whose fee schedules are set separately, were unaffected in our measurements.

Unlike the local benchmark, these deployments are publicly verifiable. Table~\ref{tab:addresses} lists the contract addresses. Through them, any reader can inspect every deployment, authorization, admitted commitment, \textsc{ModelUpdated} event, and revocation on a block explorer, along with their sender, block, and gas. The rejected attempts are the exception: they reverted during gas estimation and were never included in a block, so they appear only in our released logs. The repository also includes a script that retrieves each creation transaction from the networks and verifies that it contains exactly the released bytecode, which can be reproduced by compiling the released sources.

\begin{table}[t]
\caption{Registries deployed on public test networks in October 2026. Transactions can be inspected at \url{https://sepolia.etherscan.io} and \url{https://sepolia.arbiscan.io}.}
\label{tab:addresses}
\footnotesize
\begin{tabular}{@{}lll@{}}
\toprule
Network & Registry & Contract address \\
\midrule
Ethereum Sepolia & Allowlist (\texttt{FLRegistry}) & \texttt{0x54e698E73F26e18B17071BA471Cc6a11B6d5D57f} \\
Ethereum Sepolia & Signed credential (\texttt{FLRegistry\allowbreak ZK}) & \texttt{0x2730b56bc97560D342a2032F6C4239c56eB5667F} \\
Arbitrum Sepolia & Allowlist (\texttt{FLRegistry}) & \texttt{0xc4Ed367EEA1a699435F8351039b21e154b728476} \\
Arbitrum Sepolia & Signed credential (\texttt{FLRegistry\allowbreak ZK}) & \texttt{0x759F85E519Cb6ABcFCe14A4833ef8C071490857E} \\
\bottomrule
\end{tabular}
\end{table}

\begin{table}[t]
\caption{Registries deployed on public test networks (five clients, ten rounds per registry, October 2026), compared with the local node. Gas values are medians; A/S denotes the allowlist and signed-credential registries. Contract addresses are listed in Table~\ref{tab:addresses}.}
\label{tab:testnet}
\footnotesize\setlength{\tabcolsep}{2.5pt}
\begin{tabular}{@{}llll@{}}
\toprule
& Local node & Arbitrum Sepolia & Ethereum Sepolia \\
\midrule
Gas schedule & pre-Glamsterdam & Arbitrum & Glamsterdam \\
Deployment gas, A / S & 723{,}368 / 1{,}935{,}532 & 723{,}368 / 1{,}935{,}532 & 4{,}964{,}532 / 13{,}611{,}116 \\
Authorization gas, A / S & 46{,}333 / 80{,}134 & 46{,}333 / 80{,}134 & 128{,}353 / 250{,}062 \\
Commitment gas, A / S & 46{,}873 / 48{,}869 & 46{,}873 / 48{,}869 & 62{,}273 / 64{,}269 \\
One-transaction confirmation & 0.05 s & 1.2 s (1.0--1.6) & 10.6 s (6.8--22.5) \\
Five-client round, A / S & -- & $9.7\pm0.2$ / $9.6\pm0.3$ s & $15.7\pm7.7$ / $17.5\pm10.7$ s \\
Attack attempts accepted & 0 / 69 & 0 / 10 & 0 / 10 \\
Commitments matching (G4) & 1 / 1 per registry & 100 / 100 & 100 / 100 \\
\bottomrule
\end{tabular}
\end{table}

\begin{table}[t]
\caption{Measured gas (upper block), derived cost in US dollars and time overhead. Gas values are medians from the registry benchmark with ten clients; per-round costs use the median commitment and exclude one-off deployment and authorization. USD values assume ETH $=$ US\$3{,}000; the rollup value excludes layer-1 data fees.}
\label{tab:cost}
\footnotesize
\begin{tabular}{@{}lccc@{}}
\toprule
& Allowlist & + events & Signed VC \\
\midrule
Deployment & 723{,}368 & 759{,}286 & 1{,}935{,}532 \\
Authorization, once per client & 46{,}333 & 47{,}494 & 80{,}134 \\
Commitment, per client and round & 46{,}873 & 47{,}256 & 48{,}869 \\
Revocation & -- & -- & 52{,}724 \\
\midrule
Commitments per round, 10 clients & 468{,}730 & 472{,}560 & 488{,}690 \\
L1, 20 gwei: per round / 100 rounds & \$28.12 / \$2{,}812 & \$28.35 / \$2{,}835 & \$29.32 / \$2{,}932 \\
Arbitrum, 0.1 gwei: per round / 100 rounds & \$0.14 / \$14.06 & \$0.14 / \$14.18 & \$0.15 / \$14.66 \\
\midrule
Local training per client and round & \multicolumn{3}{c}{$5.01\pm0.07$ s} \\
Commitment latency (local node) & \multicolumn{3}{c}{$49.6\pm1.8$ ms ($0.99\%$ of training time)} \\
\bottomrule
\end{tabular}
\end{table}

\section{Discussion}\label{sec:discussion}

Every content-based defense we tested depends on an assumption about the Sybils, either that they are a minority or that they resemble one another, and an adversary that controls both the number and the content of its identities can violate either one. Identity gating removes the first lever and leaves the second to a robust rule. We answer each research question in this light and then state the limits of the conclusion.

\paragraph{RQ1: What performance does the federation reach?} The identity-gated federation reached an AUC of about 0.87 and an AUPRC about eight times the prevalence, using seven administrative variables, two of which are only known at discharge. This was within 0.007 of centralized training on the same data and above what clients reached on their own, especially for clients with few deaths. Validation-based early stopping raised it to about 0.89, at the price of worse calibration. Splitting by patient rather than by admission did not change the discrimination, so repeated patients do not explain these figures. These results are benchmark figures, not evidence of clinical readiness. Any deployment would need a prospective feature set, threshold selection for the intended use, and recalibration.

\paragraph{RQ2: how does it compare with content-based defenses?} Robust aggregation and identity gating defend different assumptions. Robust rules and norm clipping assume an honest majority and, in our experiments, worked while Sybils were a minority; there, they were also more accurate than TBFL's FedAvg. FoolsGold rests on the opposite assumption, that Sybils resemble each other more than honest clients do. It neutralized identical and label-flipping Sybils even in the majority, but it handed the entire aggregation weight to adaptive ones. Every content-based rule we tested failed against at least one strategy once Sybils made up half the participants, and FoolsGold failed against adaptive Sybils already at a third; the $\sigma$ sweep shows that an adversary aware of the defense can pick the strategy that defeats it. These failures agree with the literature rather than contradict it. The robust rules fail where their breakdown point lies~\cite{blanchard2017,yin2018}, and optimized attacks defeat them even below that point~\cite{fang2020}, so our untargeted attacks understate their weakness. FoolsGold failed exactly when its stated premise, that Sybils pursue a shared objective and therefore submit similar updates~\cite{fung2020}, no longer held. FLTrust tolerates a malicious majority, but only by giving the server a clean root dataset~\cite{cao2021fltrust}; the gate reaches the same tolerance without server-side data, at the price of trusting an issuer. Identity gating keeps the participant count tied to real, credentialed institutions, which is exactly the assumption that a Sybil adversary breaks. Its effect does not depend on what the adversary sends, and it was the same for random, adaptive, and label-flipping Sybils. The two approaches are therefore complementary, and Experiment~5 shows that the complement is needed. The gate removes the Sybil majority but admits a poisoning hospital: one or two credentialed insiders who flipped their labels drove the gate followed by FedAvg below chance in some seeds. The median applied to the admitted updates handled both threats, keeping its attack-free AUC (0.893) at every Sybil share, including the shares at which the median alone collapsed, and with one or two insiders. We therefore recommend the gate, followed by a robust rule, as the default configuration; plain FedAvg remains appropriate only when every credentialed hospital is trusted. Our insiders were few, untargeted, and honest about their sample sizes, so this recommendation rests on the breakdown point of the robust rule rather than on a guarantee. We attribute the accuracy gap between TBFL's FedAvg and the robust rules (about 0.02 in AUC) to FedAvg's drift over many rounds rather than to the gate, which admitted every honest update: the early-stopping analysis largely removed the gap ($0.889$ at the validation-selected round).

\paragraph{RQ3: when does identity gating exclude Sybils?} Under assumptions (A1)--(A4), credential-bound admission excludes every identity without a valid credential from aggregation (Proposition~\ref{prop:gate}). The deployed contracts rejected every attack case in Table~\ref{tab:sectests}, and none of the 15{,}785 Sybil commitments sent during training was admitted. The guarantee is only as strong as the issuer. A careless or corrupt issuer that issues credentials for fictitious institutions re-opens the Sybil attack. The ledger cannot prevent this, but it makes every authorization public and attributable, so such failures can be audited. Previous SSI-based FL frameworks authenticated participants and measured the cost of credential operations~\cite{geng2021didefed,haque2024ssifl,zeydan2024}; our results provide the missing link to model integrity, showing that this authentication is what keeps a Sybil majority out of the aggregate.

\paragraph{RQ4: what does on-chain admission cost?} The time overhead was about 1\% on a local node. On public test networks, a commitment was confirmed in about one second on a rollup and in about ten seconds on Ethereum Sepolia, so a layer-1 deployment must overlap commitments with training. The gas cost was constant per round and independent of model size. Verifying signed credentials on-chain added a one-off cost per credential and almost nothing per round. The monetary cost, however, depends on the ledger: about US\$28 per ten-client round on Ethereum mainnet at 20~gwei, or about US\$37 under the Glamsterdam gas schedule already active on Sepolia, but about US\$0.14 on a rollup. A health consortium would therefore deploy the registry on a rollup or on a permissioned ledger, where the cost moves to operating the infrastructure.

\paragraph{Why a ledger rather than a conventional PKI?} The same issuer could instead sign X.509 certificates~\cite{rfc5280}, and the aggregator could admit only clients that authenticate with them over mutual TLS~\cite{rfc8446}. Under (A1), this would exclude Sybils just as well, at negligible cost and latency. Table~\ref{tab:pki} compares the two designs. The difference lies not in Sybil exclusion but in who must be trusted to enforce it and who can check that it was enforced. With a PKI, admission is a decision of the aggregator's operator, and the record of which updates were accepted lives in that operator's logs. With the registry, admission is a rule executed by every ledger node, and authorizations, revocations, and commitments form a record that any consortium member can audit (G3). The registry also allows several aggregators to share a single admission state, and its commitments enable a third party to verify which updates were admitted and by whom (G4). Checking that the aggregate was computed from exactly those updates would further require verifiable aggregation or a trusted execution environment (A4). These properties matter when the hospitals do not want to trust a single operator. When one trusted operator is acceptable, a PKI is the simpler and cheaper choice. Transparency logs for certificates~\cite{rfc6962} provide some auditability without a ledger, but they do not record model commitments.

\paragraph{What the peer-to-peer ledger contributes.} The peer-to-peer dimension of TBFL lies in its admission layer, not in the training topology. Admission state is replicated across ledger nodes that no single hospital or aggregator controls; any member can verify authorizations, revocations, and commitments without trusting the aggregator; and several aggregators, or a successor to a failed one, can read the same state. Training itself still uses a single aggregator (A4), and our federated runs used a single local node, with public networks used only for the admission benchmark. Decentralizing aggregation, for example through gossip-based or multi-aggregator training, is complementary and would reuse the same registry.

\begin{table}[t]
\caption{Identity gating with a conventional PKI at the aggregator and with the ledger registry of TBFL. Both rely on the same accountable issuer (A1).}
\label{tab:pki}
\footnotesize
\begin{tabular}{@{}lll@{}}
\toprule
Property & PKI and mutual TLS & Ledger registry (TBFL) \\
\midrule
Excludes uncredentialed identities & yes, under (A1) & yes, under (A1) \\
Enforced by & aggregator operator & contract, run by every node \\
Record of admissions & operator logs & tamper-evident, public to members \\
Revocation visible to & relying party (CRL, OCSP) & every member (contract state) \\
Several aggregators & separate trust stores & shared admission state \\
Update--sender binding & needs a separate signed log & commitments on the ledger \\
Cost per round & negligible & gas or ledger operation \\
Added latency & milliseconds & ledger confirmation time \\
\bottomrule
\end{tabular}
\end{table}

\subsection{Limitations}\label{sec:limits}

\begin{enumerate}
\item \textit{Simulated federation.} The ten clients are a statistical partition of the records of one hospital system, not independent institutions with their own coding practices and case mix.
\item \textit{Repeated patients.} Partitions and splits are made by admission, so the same patient can contribute to training and test data. The patient-level sensitivity analysis found no effect on discrimination for centralized training (Table~\ref{tab:perf}), and any residual effect would apply equally to all defenses.
\item \textit{Discharge-time features.} The diagnosis and procedure counts are post-hoc summaries of the admission, which limits the clinical interpretation of the task.
\item \textit{Attack model.} We evaluate untargeted Sybil attacks: random vectors, identical or diversified, and label flipping on data from the same distribution. The random vectors are far larger than honest updates, and norm clipping rejected them when they were in the minority. The failure of the robust rules at a Sybil majority is what their breakdown point predicts. Optimized or targeted attacks~\cite{fang2020,bagdasaryan2020} would be more damaging to content-based defenses, but not to the gate, whose effect does not depend on update content. Credentialed insiders were tested only as one or two hospitals that flipped their own labels (Experiment~5). We did not test colluding or targeted insiders, or insiders that misreport their sample sizes to gain FedAvg weight; the gate offers no protection against any of these (Section~\ref{sec:threatmodel}).
\item \textit{Statistical power.} The main conditions use five seeds; norm clipping, the gate with the median, the smaller $\sigma$ values, Experiments~4 and~5, and the gate with 15 adaptive Sybils use three. The large effects are clear, but small differences among robust aggregators with fewer than 50\% Sybils are not resolved.
\item \textit{Ledger.} The federated training ran against a local node; the registries were deployed on public test networks only for the admission benchmark, with five clients and from a single machine. Zero-knowledge membership proofs were released but not evaluated (their nullifiers are bound to a counter of accepted submissions, not to FL rounds), mainnet costs were extrapolated from measured gas, and permissioned ledgers were not deployed.
\end{enumerate}

\section{Conclusion}\label{sec:conclusion}

For federations whose members are licensed institutions, the most effective defense against Sybil majorities that we found is not a better aggregation statistic but an accountable answer to the question of who may participate.

Content-based defenses protected the model only under the conditions they assume: a minority of Sybils for robust aggregation and norm clipping, and mutually similar Sybils for FoolsGold. An adversary that controls both the number and the content of its identities can violate either condition. Binding participation to accountable institutional identity removes the first of these levers, at about one percent of training time on a local node and a ledger cost that a rollup makes affordable, and leaves the second to a robust rule applied to the admitted updates. Identity gating does not protect against compromised credentialed hospitals: a single one that flipped its labels lowered the mean AUC of plain averaging over three seeds from 0.872 to 0.746, with one seed falling from 0.830 to 0.480. It should therefore be combined with robust aggregation rather than replaced; in our experiments, the combination withstood both Sybil majorities and credentialed insiders at the accuracy of the robust rule alone.

Future work will run federated training against public and permissioned networks, measure zero-knowledge presentations, evaluate the gate combined with robust aggregation against colluding, targeted and optimized insider attacks and misreported sample sizes, and validate the approach on a multi-institutional cohort with prospective features.

\section*{Acknowledgements}
The authors thank the MIT Laboratory for Computational Physiology for maintaining MIMIC-IV.

\section*{Statements and Declarations}

\begin{itemize}
\item \textbf{Funding:} The authors declare that no funds, grants, or other support were received during the preparation of this manuscript.
\item \textbf{Competing interests:} The authors have no relevant financial or non-financial interests to disclose.
\item \textbf{Ethics approval:} This study used the de-identified MIMIC-IV database under the PhysioNet Credentialed Health Data Use Agreement. No new data were collected from human subjects.
\item \textbf{Data availability:} MIMIC-IV is available from PhysioNet (\url{https://physionet.org/content/mimiciv/}) to credentialed users. Under the data use agreement, the authors do not redistribute MIMIC-IV data or derived patient-level files.
\item \textbf{Code availability:} The SQL cohort definition, the code for partitioning the data, running the experiments, deploying and benchmarking the contracts and producing the figures, per-run logs of metrics and transactions without patient-level data, and the addresses of the contracts deployed on the public test networks are archived on Zenodo, and the version used in this paper (v2.1) is available at \url{https://doi.org/10.5281/zenodo.23836753}~\cite{tbfl_code_v21}. Development continues at \url{https://github.com/rodrigoronner/TBFL-EHR-Framework}.
\item \textbf{Author contributions:} R.T., R.A.\ and L.A.\ conceived and designed the study. R.T.\ implemented the framework, processed the data, and ran the experiments. R.A.\ and L.A.\ supervised the project and critically revised the manuscript. All authors approved the final version.
\end{itemize}

\bibliographystyle{unsrtnat}
\bibliography{references}

\begin{thebibliography}{61}
\providecommand{\natexlab}[1]{#1}
\providecommand{\url}[1]{\texttt{#1}}
\expandafter\ifx\csname urlstyle\endcsname\relax
  \providecommand{\doi}[1]{doi: #1}\else
  \providecommand{\doi}{doi: \begingroup \urlstyle{rm}\Url}\fi

\bibitem[Shah and Khan(2020)]{shah2020}
Syed~Mohsin Shah and Rafiq~Ahmad Khan.
\newblock Secondary use of electronic health record: Opportunities and
  challenges.
\newblock \emph{IEEE Access}, 8:\penalty0 136947--136965, 2020.
\newblock \doi{10.1109/ACCESS.2020.3011099}.

\bibitem[McMahan et~al.(2017)McMahan, Moore, Ramage, Hampson, and
  Arcas]{mcmahan2017}
Brendan McMahan, Eider Moore, Daniel Ramage, Seth Hampson, and Blaise
  Ag{\"u}era~y Arcas.
\newblock Communication-efficient learning of deep networks from decentralized
  data.
\newblock In \emph{Proceedings of the 20th International Conference on
  Artificial Intelligence and Statistics (AISTATS)}, volume~54 of \emph{PMLR},
  pages 1273--1282, 2017.

\bibitem[Rieke et~al.(2020)Rieke, Hancox, Li, et~al.]{rieke2020}
Nicola Rieke, Jonny Hancox, Wenqi Li, et~al.
\newblock The future of digital health with federated learning.
\newblock \emph{npj Digital Medicine}, 3:\penalty0 119, 2020.
\newblock \doi{10.1038/s41746-020-00323-1}.

\bibitem[Kaissis et~al.(2020)Kaissis, Makowski, R{\"u}ckert, and
  Braren]{kaissis2020}
Georgios~A. Kaissis, Marcus~R. Makowski, Daniel R{\"u}ckert, and Rickmer~F.
  Braren.
\newblock Secure, privacy-preserving and federated machine learning in medical
  imaging.
\newblock \emph{Nature Machine Intelligence}, 2\penalty0 (6):\penalty0
  305--311, 2020.
\newblock \doi{10.1038/s42256-020-0186-1}.

\bibitem[Douceur(2002)]{douceur2002}
John~R. Douceur.
\newblock The {S}ybil attack.
\newblock In \emph{Peer-to-Peer Systems (IPTPS 2002)}, volume 2429 of
  \emph{Lecture Notes in Computer Science}, pages 251--260, Berlin, Heidelberg,
  2002. Springer.
\newblock \doi{10.1007/3-540-45748-8_24}.

\bibitem[Fung et~al.(2020)Fung, Yoon, and Beschastnikh]{fung2020}
Clement Fung, Chris J.~M. Yoon, and Ivan Beschastnikh.
\newblock The limitations of federated learning in {S}ybil settings.
\newblock In \emph{23rd International Symposium on Research in Attacks,
  Intrusions and Defenses (RAID 2020)}, pages 301--316, 2020.

\bibitem[Lian et~al.(2023)Lian, Zhang, Nan, and Su]{lian2023}
Zhuotao Lian, Chen Zhang, Kai Nan, and Chunhua Su.
\newblock {SPoiL}: {S}ybil-based untargeted data poisoning attacks in federated
  learning.
\newblock In \emph{Network and System Security (NSS 2023)}, Lecture Notes in
  Computer Science, pages 235--248, Cham, 2023. Springer.
\newblock \doi{10.1007/978-3-031-39828-5_13}.

\bibitem[Samy and Girdzijauskas(2023)]{samy2023}
Ahmed~E. Samy and {\v{S}}ar{\=u}nas Girdzijauskas.
\newblock Mitigating {S}ybil attacks in federated learning.
\newblock In \emph{Information Security Practice and Experience (ISPEC 2023)},
  Lecture Notes in Computer Science, pages 36--51, Singapore, 2023. Springer.
\newblock \doi{10.1007/978-981-99-7032-2_3}.

\bibitem[Blanchard et~al.(2017)Blanchard, El~Mhamdi, Guerraoui, and
  Stainer]{blanchard2017}
Peva Blanchard, El~Mahdi El~Mhamdi, Rachid Guerraoui, and Julien Stainer.
\newblock Machine learning with adversaries: {B}yzantine tolerant gradient
  descent.
\newblock In \emph{Advances in Neural Information Processing Systems 30
  (NeurIPS 2017)}, pages 119--129, 2017.

\bibitem[Yin et~al.(2018)Yin, Chen, Kannan, and Bartlett]{yin2018}
Dong Yin, Yudong Chen, Ramchandran Kannan, and Peter Bartlett.
\newblock {B}yzantine-robust distributed learning: Towards optimal statistical
  rates.
\newblock In \emph{Proceedings of the 35th International Conference on Machine
  Learning (ICML)}, volume~80 of \emph{PMLR}, pages 5650--5659, 2018.

\bibitem[Kim et~al.(2020)Kim, Park, Bennis, and Kim]{kim2020blockfl}
Hyesung Kim, Jihong Park, Mehdi Bennis, and Seong-Lyun Kim.
\newblock Blockchained on-device federated learning.
\newblock \emph{IEEE Communications Letters}, 24\penalty0 (6):\penalty0
  1279--1283, 2020.
\newblock \doi{10.1109/LCOMM.2019.2921755}.

\bibitem[Lu et~al.(2020)Lu, Huang, Dai, Maharjan, and Zhang]{lu2020}
Yunlong Lu, Xiaohong Huang, Yueyue Dai, Sabita Maharjan, and Yan Zhang.
\newblock Blockchain and federated learning for privacy-preserved data sharing
  in industrial {IoT}.
\newblock \emph{IEEE Transactions on Industrial Informatics}, 16\penalty0
  (6):\penalty0 4177--4186, 2020.
\newblock \doi{10.1109/TII.2019.2942190}.

\bibitem[Weng et~al.(2021)Weng, Weng, Zhang, Li, Zhang, and Luo]{weng2021}
Jiasi Weng, Jian Weng, Jilian Zhang, Ming Li, Yue Zhang, and Weiqi Luo.
\newblock {DeepChain}: Auditable and privacy-preserving deep learning with
  blockchain-based incentive.
\newblock \emph{IEEE Transactions on Dependable and Secure Computing},
  18\penalty0 (5):\penalty0 2438--2455, 2021.
\newblock \doi{10.1109/TDSC.2019.2952332}.

\bibitem[Qammar et~al.(2023)Qammar, Karim, Ning, and Ding]{qammar2023}
Attia Qammar, Ahmad Karim, Huansheng Ning, and Jianguo Ding.
\newblock Securing federated learning with blockchain: A systematic literature
  review.
\newblock \emph{Artificial Intelligence Review}, 56\penalty0 (5):\penalty0
  3951--3985, 2023.
\newblock \doi{10.1007/s10462-022-10271-9}.

\bibitem[Sporny et~al.(2022)Sporny, Guy, Sabadello, and Reed]{w3c_did}
Manu Sporny, Amy Guy, Markus Sabadello, and Drummond Reed.
\newblock Decentralized identifiers ({DIDs}) v1.0: Core architecture, data
  model, and representations.
\newblock W3c recommendation, World Wide Web Consortium (W3C), 2022.
\newblock URL \url{https://www.w3.org/TR/2022/REC-did-core-20220719/}.

\bibitem[Sporny et~al.(2019)Sporny, Longley, and Chadwick]{w3c_vc}
Manu Sporny, Dave Longley, and David Chadwick.
\newblock Verifiable credentials data model 1.0: Expressing verifiable
  information on the web.
\newblock W3c recommendation, World Wide Web Consortium (W3C), 2019.
\newblock URL \url{https://www.w3.org/TR/2019/REC-vc-data-model-20191119/}.

\bibitem[Houtan et~al.(2020)Houtan, Hafid, and Makrakis]{houtan2020}
Bahar Houtan, Abdelhakim~Senhaji Hafid, and Dimitrios Makrakis.
\newblock A survey on blockchain-based self-sovereign patient identity in
  healthcare.
\newblock \emph{IEEE Access}, 8:\penalty0 90478--90494, 2020.
\newblock \doi{10.1109/ACCESS.2020.2994090}.

\bibitem[Cocco and Tonelli(2024)]{cocco2024}
Luisanna Cocco and Roberto Tonelli.
\newblock A self-sovereign identity--blockchain-based model proposal for deep
  digital transformation in the healthcare sector.
\newblock \emph{Future Internet}, 16\penalty0 (12):\penalty0 473, 2024.
\newblock \doi{10.3390/fi16120473}.

\bibitem[L{\'o}pez~Mart{\'i}nez et~al.(2026)L{\'o}pez~Mart{\'i}nez, Naghmouchi,
  Laurent, et~al.]{martinez2026}
Antonio L{\'o}pez~Mart{\'i}nez, Montassar Naghmouchi, Maryline Laurent, et~al.
\newblock Breaking barriers in healthcare: A secure identity framework for
  seamless access.
\newblock \emph{Computer Standards \& Interfaces}, 95:\penalty0 104020, 2026.
\newblock \doi{10.1016/j.csi.2025.104020}.

\bibitem[Geng et~al.(2021)Geng, Kanwal, Jaatun, and Rong]{geng2021didefed}
Jiahui Geng, Neel Kanwal, Martin~Gilje Jaatun, and Chunming Rong.
\newblock {DID-eFed}: Facilitating federated learning as a service with
  decentralized identities.
\newblock In \emph{Evaluation and Assessment in Software Engineering (EASE
  2021)}, pages 329--335, New York, NY, USA, 2021. ACM.
\newblock \doi{10.1145/3463274.3463352}.

\bibitem[Haque et~al.(2022)Haque, Hasan, Daria, Qu, and Jiang]{haque2022}
Rakib~Ul Haque, A.~S. M.~Touhidul Hasan, Apubra Daria, Qiang Qu, and Qingshan
  Jiang.
\newblock Towards convergence of blockchain and self-sovereign identity for
  privacy-preserving secure federated learning.
\newblock In \emph{Big Data and Security (ICBDS 2021)}, Communications in
  Computer and Information Science, pages 243--255, Singapore, 2022. Springer.
\newblock \doi{10.1007/978-981-19-0852-1_19}.

\bibitem[Haque et~al.(2024)Haque, Hasan, Al-Hababi, Zhang, and
  Xu]{haque2024ssifl}
Rakib~Ul Haque, A.~S. M.~Touhidul Hasan, Mohammed Ali~Mohammed Al-Hababi,
  Yuqing Zhang, and Dianxiang Xu.
\newblock {SSI-FL}: Self-sovereign identity based privacy-preserving federated
  learning.
\newblock \emph{Journal of Parallel and Distributed Computing}, 191:\penalty0
  104907, 2024.
\newblock \doi{10.1016/j.jpdc.2024.104907}.

\bibitem[Zeydan et~al.(2023)Zeydan, Blanco, Mangues, Arslan, and
  Turk]{zeydan2023}
Engin Zeydan, Luis Blanco, Josep Mangues, Suayb Arslan, and Yekta Turk.
\newblock Blockchain-based self-sovereign identity for federated learning in
  vehicular networks.
\newblock In \emph{19th International Conference on Network and Service
  Management (CNSM 2023)}, pages 1--7, 2023.
\newblock \doi{10.23919/CNSM59352.2023.10327817}.

\bibitem[Zeydan et~al.(2024)Zeydan, Blanco, Mangues-Bafalluy, Arslan, Turk,
  Yadav, and Liyanage]{zeydan2024}
Engin Zeydan, Luis Blanco, Josep Mangues-Bafalluy, Suayb~S. Arslan, Yekta Turk,
  Awaneesh~Kumar Yadav, and Madhusanka Liyanage.
\newblock Blockchain-based self-sovereign identity: Taking control of identity
  in federated learning.
\newblock \emph{IEEE Open Journal of the Communications Society}, 5:\penalty0
  5764--5781, 2024.
\newblock \doi{10.1109/OJCOMS.2024.3449692}.

\bibitem[Cao et~al.(2021)Cao, Fang, Liu, and Gong]{cao2021fltrust}
Xiaoyu Cao, Minghong Fang, Jia Liu, and Neil~Zhenqiang Gong.
\newblock {FLTrust}: {B}yzantine-robust federated learning via trust
  bootstrapping.
\newblock In \emph{Proceedings of the Network and Distributed System Security
  Symposium (NDSS 2021)}, 2021.
\newblock \doi{10.14722/ndss.2021.24434}.

\bibitem[Bagdasaryan et~al.(2020)Bagdasaryan, Veit, Hua, Estrin, and
  Shmatikov]{bagdasaryan2020}
Eugene Bagdasaryan, Andreas Veit, Yiqing Hua, Deborah Estrin, and Vitaly
  Shmatikov.
\newblock How to backdoor federated learning.
\newblock In \emph{Proceedings of the 23rd International Conference on
  Artificial Intelligence and Statistics (AISTATS)}, volume 108 of \emph{PMLR},
  pages 2938--2948, 2020.

\bibitem[Fang et~al.(2020)Fang, Cao, Jia, and Gong]{fang2020}
Minghong Fang, Xiaoyu Cao, Jinyuan Jia, and Neil Gong.
\newblock Local model poisoning attacks to {B}yzantine-robust federated
  learning.
\newblock In \emph{29th USENIX Security Symposium}, pages 1605--1622, 2020.

\bibitem[Fang et~al.(2024)Fang, Zhang, Hairi, Khanduri, Liu, Lu, Liu, and
  Gong]{fang2024}
Minghong Fang, Zifan Zhang, Hairi, Prashant Khanduri, Jia Liu, Songtao Lu,
  Yuchen Liu, and Neil Gong.
\newblock {B}yzantine-robust decentralized federated learning.
\newblock In \emph{Proceedings of the 2024 ACM SIGSAC Conference on Computer
  and Communications Security (CCS '24)}, pages 2874--2888, 2024.
\newblock \doi{10.1145/3658644.3670307}.

\bibitem[Cajaraville-Aboy et~al.(2025)Cajaraville-Aboy, Fern{\'a}ndez-Vilas,
  D{\'i}az-Redondo, et~al.]{cajaraville2025}
Diego Cajaraville-Aboy, Ana Fern{\'a}ndez-Vilas, Rebeca~P. D{\'i}az-Redondo,
  et~al.
\newblock {B}yzantine-robust aggregation for securing decentralized federated
  learning.
\newblock \emph{IEEE Access}, 13:\penalty0 190947--190963, 2025.
\newblock \doi{10.1109/ACCESS.2025.3629864}.

\bibitem[Wang et~al.(2025)Wang, Zheng, and Lin]{wang2025trimmed}
Tianxiang Wang, Zhonglong Zheng, and Feilong Lin.
\newblock Federated learning framework based on trimmed mean aggregation rules.
\newblock \emph{Expert Systems with Applications}, 270:\penalty0 126354, 2025.
\newblock \doi{10.1016/j.eswa.2024.126354}.

\bibitem[Krau{\ss} and Dmitrienko(2023)]{krauss2023}
Torsten Krau{\ss} and Alexandra Dmitrienko.
\newblock {MESAS}: Poisoning defense for federated learning resilient against
  adaptive attackers.
\newblock In \emph{Proceedings of the 2023 ACM SIGSAC Conference on Computer
  and Communications Security (CCS '23)}, pages 1526--1540, 2023.
\newblock \doi{10.1145/3576915.3623212}.

\bibitem[Zhu et~al.(2019)Zhu, Liu, and Han]{zhu2019dlg}
Ligeng Zhu, Zhijian Liu, and Song Han.
\newblock Deep leakage from gradients.
\newblock In \emph{Advances in Neural Information Processing Systems 32
  (NeurIPS 2019)}, 2019.

\bibitem[Bonawitz et~al.(2017)Bonawitz, Ivanov, Kreuter, et~al.]{bonawitz2017}
Keith Bonawitz, Vladimir Ivanov, Ben Kreuter, et~al.
\newblock Practical secure aggregation for privacy-preserving machine learning.
\newblock In \emph{Proceedings of the 2017 ACM SIGSAC Conference on Computer
  and Communications Security (CCS '17)}, pages 1175--1191, 2017.
\newblock \doi{10.1145/3133956.3133982}.

\bibitem[Abadi et~al.(2016)Abadi, Chu, Goodfellow, McMahan, Mironov, Talwar,
  and Zhang]{abadi2016}
Mart{\'i}n Abadi, Andy Chu, Ian Goodfellow, H.~Brendan McMahan, Ilya Mironov,
  Kunal Talwar, and Li~Zhang.
\newblock Deep learning with differential privacy.
\newblock In \emph{Proceedings of the 2016 ACM SIGSAC Conference on Computer
  and Communications Security (CCS '16)}, pages 308--318, 2016.
\newblock \doi{10.1145/2976749.2978318}.

\bibitem[Nguyen et~al.(2021)Nguyen, Ding, Pham, Pathirana, Le, Seneviratne, Li,
  Niyato, and Poor]{nguyen2021}
Dinh~C. Nguyen, Ming Ding, Quoc-Viet Pham, Pubudu~N. Pathirana, Long~Bao Le,
  Aruna Seneviratne, Jun Li, Dusit Niyato, and H.~Vincent Poor.
\newblock Federated learning meets blockchain in edge computing: Opportunities
  and challenges.
\newblock \emph{IEEE Internet of Things Journal}, 8\penalty0 (16):\penalty0
  12806--12825, 2021.
\newblock \doi{10.1109/JIOT.2021.3072611}.

\bibitem[Yang et~al.(2023)Yang, Shi, Zhou, Wang, and Yang]{yang2023tfl}
Zhanpeng Yang, Yuanming Shi, Yong Zhou, Zixin Wang, and Kai Yang.
\newblock Trustworthy federated learning via blockchain.
\newblock \emph{IEEE Internet of Things Journal}, 10\penalty0 (1):\penalty0
  92--109, 2023.
\newblock \doi{10.1109/JIOT.2022.3201117}.

\bibitem[Ning et~al.(2024)Ning, Zhu, Song, et~al.]{ning2024}
Weiguang Ning, Yingjuan Zhu, Caixia Song, et~al.
\newblock Blockchain-based federated learning: A survey and new perspectives.
\newblock \emph{Applied Sciences}, 14\penalty0 (20):\penalty0 9459, 2024.
\newblock \doi{10.3390/app14209459}.

\bibitem[Shawkat et~al.(2025)Shawkat, El-desoky, Ali, and Salem]{shawkat2025}
Mai Shawkat, Ali El-desoky, Zainab~H. Ali, and Mofreh Salem.
\newblock Blockchain and federated learning based on aggregation techniques for
  industrial {IoT}: A contemporary survey.
\newblock \emph{Peer-to-Peer Networking and Applications}, 18\penalty0
  (4):\penalty0 192, 2025.
\newblock \doi{10.1007/s12083-025-01991-0}.

\bibitem[Wang et~al.(2022)Wang, Yan, and Dong]{wang2022shapley}
Zhe Wang, Bo~Yan, and Aixin Dong.
\newblock Blockchain empowered federated learning for data sharing incentive
  mechanism.
\newblock \emph{Procedia Computer Science}, 202:\penalty0 348--353, 2022.
\newblock \doi{10.1016/j.procs.2022.04.047}.

\bibitem[Lo et~al.(2023)Lo, Liu, Lu, Wang, Xu, Paik, and Zhu]{lo2023}
Sin~Kit Lo, Yue Liu, Qinghua Lu, Chen Wang, Xiwei Xu, Hye-Young Paik, and
  Liming Zhu.
\newblock Toward trustworthy {AI}: Blockchain-based architecture design for
  accountability and fairness of federated learning systems.
\newblock \emph{IEEE Internet of Things Journal}, 10\penalty0 (4):\penalty0
  3276--3284, 2023.
\newblock \doi{10.1109/JIOT.2022.3144450}.

\bibitem[Hasselgren et~al.(2020)Hasselgren, Kralevska, Gligoroski, Pedersen,
  and Faxvaag]{hasselgren2020}
Anton Hasselgren, Katina Kralevska, Danilo Gligoroski, Sindre~A. Pedersen, and
  Arild Faxvaag.
\newblock Blockchain in healthcare and health sciences---a scoping review.
\newblock \emph{International Journal of Medical Informatics}, 134:\penalty0
  104040, 2020.
\newblock \doi{10.1016/j.ijmedinf.2019.104040}.

\bibitem[H{\"o}lbl et~al.(2018)H{\"o}lbl, Kompara, Kami{\v{s}}ali{\'c}, and
  Nemec~Zlatolas]{holbl2018}
Marko H{\"o}lbl, Marko Kompara, Aida Kami{\v{s}}ali{\'c}, and Lili
  Nemec~Zlatolas.
\newblock A systematic review of the use of blockchain in healthcare.
\newblock \emph{Symmetry}, 10\penalty0 (10):\penalty0 470, 2018.
\newblock \doi{10.3390/sym10100470}.

\bibitem[Naghmouchi and Laurent(2025)]{naghmouchi2025}
Montassar Naghmouchi and Maryline Laurent.
\newblock A systematic review and layered framework for privacy-by-design in
  self-sovereign identity systems.
\newblock arXiv preprint arXiv:2502.02520, 2025.

\bibitem[Li et~al.(2020)Li, Sahu, Zaheer, Sanjabi, Talwalkar, and
  Smith]{li2020fedprox}
Tian Li, Anit~Kumar Sahu, Manzil Zaheer, Maziar Sanjabi, Ameet Talwalkar, and
  Virginia Smith.
\newblock Federated optimization in heterogeneous networks.
\newblock In \emph{Proceedings of Machine Learning and Systems}, volume~2,
  pages 429--450, 2020.

\bibitem[Wood(2014)]{wood2014}
Gavin Wood.
\newblock Ethereum: A secure decentralised generalised transaction ledger.
\newblock Ethereum Project Yellow Paper, 2014.

\bibitem[Kalodner et~al.(2018)Kalodner, Goldfeder, Chen, Weinberg, and
  Felten]{kalodner2018}
Harry Kalodner, Steven Goldfeder, Xiaoqi Chen, S.~Matthew Weinberg, and
  Edward~W. Felten.
\newblock Arbitrum: Scalable, private smart contracts.
\newblock In \emph{27th USENIX Security Symposium}, pages 1353--1370, 2018.

\bibitem[Androulaki et~al.(2018)Androulaki, Barger, Bortnikov,
  et~al.]{androulaki2018}
Elli Androulaki, Artem Barger, Vita Bortnikov, et~al.
\newblock Hyperledger fabric: A distributed operating system for permissioned
  blockchains.
\newblock In \emph{Proceedings of the Thirteenth EuroSys Conference}, pages
  1--15, New York, NY, USA, 2018. ACM.
\newblock \doi{10.1145/3190508.3190538}.
\newblock Article 30.

\bibitem[{National Institute of Standards and Technology}(2015)]{fips180}
{National Institute of Standards and Technology}.
\newblock Secure hash standard ({SHS}).
\newblock Federal Information Processing Standards Publication FIPS 180-4,
  NIST, 2015.

\bibitem[Benet(2014)]{benet2014}
Juan Benet.
\newblock {IPFS} -- content addressed, versioned, {P2P} file system.
\newblock arXiv preprint arXiv:1407.3561, 2014.

\bibitem[Johnson et~al.(2023)Johnson, Bulgarelli, Shen,
  et~al.]{johnson2023mimic}
Alistair E.~W. Johnson, Lucas Bulgarelli, Lu~Shen, et~al.
\newblock {MIMIC-IV}, a freely accessible electronic health record dataset.
\newblock \emph{Scientific Data}, 10:\penalty0 1, 2023.
\newblock \doi{10.1038/s41597-022-01899-x}.

\bibitem[Goldberger et~al.(2000)Goldberger, Amaral, Glass,
  et~al.]{goldberger2000}
Ary~L. Goldberger, Luis A.~N. Amaral, Leon Glass, et~al.
\newblock {PhysioBank}, {PhysioToolkit}, and {PhysioNet}: Components of a new
  research resource for complex physiologic signals.
\newblock \emph{Circulation}, 101\penalty0 (23):\penalty0 e215--e220, 2000.
\newblock \doi{10.1161/01.CIR.101.23.e215}.

\bibitem[Collins et~al.(2015)Collins, Reitsma, Altman, and
  Moons]{collins2015tripod}
Gary~S. Collins, Johannes~B. Reitsma, Douglas~G. Altman, and Karel G.~M. Moons.
\newblock Transparent reporting of a multivariable prediction model for
  individual prognosis or diagnosis ({TRIPOD}): The {TRIPOD} statement.
\newblock \emph{BMJ}, 350:\penalty0 g7594, 2015.
\newblock \doi{10.1136/bmj.g7594}.

\bibitem[Hsu et~al.(2019)Hsu, Qi, and Brown]{hsu2019}
Tzu-Ming~Harry Hsu, Hang Qi, and Matthew Brown.
\newblock Measuring the effects of non-identical data distribution for
  federated visual classification.
\newblock arXiv preprint arXiv:1909.06335, 2019.

\bibitem[Batista et~al.(2004)Batista, Prati, and Monard]{batista2004}
Gustavo E. A. P.~A. Batista, Ronaldo~C. Prati, and Maria~Carolina Monard.
\newblock A study of the behavior of several methods for balancing machine
  learning training data.
\newblock \emph{ACM SIGKDD Explorations Newsletter}, 6\penalty0 (1):\penalty0
  20--29, 2004.
\newblock \doi{10.1145/1007730.1007735}.

\bibitem[Saito and Rehmsmeier(2015)]{saito2015}
Takaya Saito and Marc Rehmsmeier.
\newblock The precision-recall plot is more informative than the {ROC} plot
  when evaluating binary classifiers on imbalanced datasets.
\newblock \emph{PLoS ONE}, 10\penalty0 (3):\penalty0 e0118432, 2015.
\newblock \doi{10.1371/journal.pone.0118432}.

\bibitem[Van~Calster et~al.(2019)Van~Calster, McLernon, van Smeden, Wynants,
  and Steyerberg]{vancalster2019}
Ben Van~Calster, David~J. McLernon, Maarten van Smeden, Laure Wynants, and
  Ewout~W. Steyerberg.
\newblock Calibration: The {A}chilles heel of predictive analytics.
\newblock \emph{BMC Medicine}, 17:\penalty0 230, 2019.
\newblock \doi{10.1186/s12916-019-1466-7}.

\bibitem[{Ethereum Foundation}(2026)]{ef2026glamsterdam}
{Ethereum Foundation}.
\newblock Glamsterdam testnet announcement.
\newblock Ethereum Foundation Blog, September 2026.
\newblock URL
  \url{https://blog.ethereum.org/2026/09/17/glamsterdam-testnet-announcement}.

\bibitem[Cooper et~al.(2008)Cooper, Santesson, Farrell, Boeyen, Housley, and
  Polk]{rfc5280}
David Cooper, Stefan Santesson, Stephen Farrell, Sharon Boeyen, Russell
  Housley, and Tim Polk.
\newblock Internet {X.509} public key infrastructure certificate and
  certificate revocation list ({CRL}) profile.
\newblock RFC 5280, Internet Engineering Task Force, 2008.

\bibitem[Rescorla(2018)]{rfc8446}
Eric Rescorla.
\newblock The transport layer security ({TLS}) protocol version 1.3.
\newblock RFC 8446, Internet Engineering Task Force, 2018.

\bibitem[Laurie et~al.(2013)Laurie, Langley, and Kasper]{rfc6962}
Ben Laurie, Adam Langley, and Emilia Kasper.
\newblock Certificate transparency.
\newblock RFC 6962, Internet Engineering Task Force, 2013.

\bibitem[Tertulino et~al.(2026)Tertulino, Almeida, and Alencar]{tbfl_code_v21}
Rodrigo Tertulino, Ricardo Almeida, and Laercio Alencar.
\newblock {TBFL-EHR-Framework}: identity-gated federated learning for
  electronic health records (code, contracts and run logs), version 2.1.
\newblock Zenodo, 2026.
\newblock URL \url{https://doi.org/10.5281/zenodo.23836753}.

\end{thebibliography}

\end{document}